\documentclass[amsmath, amssymb,10pt, aps, prb, twocolumn, notitlepage, eqsecnum, longbibliography, superscriptaddress]{revtex4-1}

\usepackage{graphicx}
\usepackage{calc}
\usepackage{braket}
\usepackage{ulem}   
\usepackage{slashed}
\usepackage{wasysym}
\usepackage{dsfont}
\usepackage{amsthm,amsmath,amsfonts,amssymb,verbatim,color}
\usepackage{lipsum}
\usepackage{bm}
\usepackage{epsfig,slashed}
\usepackage{hyperref}
\usepackage{flafter}
\usepackage{mathtools}

\newcommand{\bA}{{\bf A}}

\newcommand{\bk}{{\bf k}}
\newcommand{\bK}{{\bf K}}

\newcommand{\bp}{{\bf p}}
\newcommand{\bR}{{\bf R}}
\newcommand{\br}{{\bf r}}
\newcommand{\bB}{{\bf B}}

\newcommand{\bq}{{\bf q}}
\newcommand{\bv}{{\bf v}}

\def\bnabla{{\bm \nabla}}
\def\hbsigma{\hat{\boldsymbol \sigma}}

\newcommand{\ha}{{\hat a}}

\newcommand{\hsigma}{{\hat\sigma}}

\newcommand{\cB}{{\cal B}}

\newcommand{\cH}{\hat{\cal H}}

\newcommand{\DisAverage}[1]{\left< #1 \right>_{\text{dis}}}

\begin{document}

\title{Magnetotransport and internodal tunnelling in Weyl semimetals}

\author{G.~Bednik}
\affiliation{Physics Department, University of California, Santa Cruz, California 95064, USA}

\author{K.S.~Tikhonov}
\affiliation{L. D. Landau Institute for Theoretical Physics, 119334 Moscow, Russia}
\affiliation{Condensed-matter Physics Laboratory, National Research University Higher School of Economics, 101000, Moscow, Russia}

\author{S.V.~Syzranov}
\affiliation{Physics Department, University of California, Santa Cruz, California 95064, USA}

\begin{abstract}
	Internodal dynamics of quasiparticles in Weyl semimetals manifest themselves in hydrodynamic, transport and thermodynamic phenomena and
	are essential for potential valleytronic applications of these systems. 
	In an external magnetic field, 
	coherent quasiparticle tunnelling between the nodes
	modifies the quasiparticle dispersion
	and, in particular, opens gaps in the dispersion of quasiparticles at the zeroth Landau level.
	We study magnetotransport in a Weyl semimetal taking into account mechanisms of quasiparticle scattering 
	both affected by such gaps and independent of them.
	We compute the longitudal resistivity of a disordered Weyl semimetal with two nodes
	in a strong magnetic field microscopically
	 and demonstrate that in a broad range of magnetic fields it has a 
	strong angular dependence $\rho(\eta)\propto C_1+C_2 \cos^2\eta$, where $\eta$ is the angle between the field
	and the separation between the nodes in momentum space. The first term is determined by the coherent internodal
	tunnelling and is important only at angles $\eta$ close to $\pi/2$. This contribution depends exponentially on the magnetic 
	field, $\propto \exp\left(-B_0/B\right)$. The second term is weakly dependent on the absolute value of the magnetic
	field for realistic concentrations of the impurities in a broad interval of fields.
\end{abstract}


\maketitle


\section{Introduction}

Recent prediction~\cite{WanVishwanath:WeylFirst} and experimental discovery~\cite{Xu:TaAs, Lv:discovery, Lv:2015, Xu:WeylDiscovery, Xu:NbAs, Yang:TaAs, Xu:TaP, Xu:TaP2, ZHasan:reviewDiscovery} of quasiparticles with Weyl dispersion in solid-state systems
have motivated a vast number of predictions and observations of novel fundamental effects involving Weyl particles (see, e.g., Refs.~\onlinecite{Armitage:WeylReview}
and \onlinecite{Vafek:DiracWeylReview} for a review).
Some of these phenomena, such as the chiral anomaly~\cite{Burkov:review,Burkov:IOPcmeReview,SonSpivak:NMR}, 
rely on the transfer of quasiparticles between different Weyl nodes, sometimes
also referred to as valleys, points in momentum space in whose
vicinity the quasiparticles display Weyl dispersion.

Weyl nodes in solid-state systems come in pairs with opposite chiralities~\cite{NielsenNinomiya}.
Apart from fundamental interest, internodal dynamics of Weyl quasiparticles may be used for valleytronic applications, i.e. using 
the valley degree of freedom
to store, process and transfer information and to control electron transport~\cite{Kharzeev:valleytronics}. 
Experimentally observed manifestations of the internodal (intervalley) dynamics
include also negative longitudinal magnetoresistnace (see, e.g., Refs.~\onlinecite{SonSpivak:NMR,Huang:negativeMR,ZHasan:AnomalySignatures,Yang:negativeMR, JianHua:negativeMR,Wang:negativeMR,Jia:ReviewMR, Xiong:evidence, Hirschberger:chiralanomaly,Ong:anomalyExperiment}), a consequence of the chiral anomaly~\cite{Burkov:review,NielsenNinomiya},
and quantum oscillations of resistance in thin slabs of Weyl semimetals~\cite{Moll:SlabOscillations,Potter:OscillationsPrediction} (WSMs).
Changing the valley degree of freedom of quasiparticles has also been predicted to lead to the release or absorption of
heat (``adiabatic dechiralisation''~\cite{Syzranov:dechiralisation}) and to affect the hydrodynamic flows of electrons in WSMs~\cite{Gorbar:WeylHydro,Gorbar:hydroCS,LucasSachdev:WeylHydrodynamics,Yamamoto:HEPhydrodynamics,Galitski:dynamo,Gorbar:hydroNonLocal,Sukhachov:hydroCollective,Gooth:hydroObservation}.

Of fundamental importance for valleytronic applications is coherent tunnelling between Weyl nodes.
Such tunnelling leads to an effective coupling between the states of quasiparticles near different nodes, which may be controlled 
by the direction and the magnitude of an external magnetic field.
This coupling leads to the opening of a gap in the quasiparticle dispersion at the zeroth Landau level in WSMs~\cite{Kim:firstGap,ChanLee:gapping,SaykinTikhonovRodionov},
which has recently been observed in experiment~\cite{HsuZhasanHsinLin:LLs,Jiang:GapObservation,ZhangHasan:gapObservation}.

In this paper, we study the interplay of magnetotransport in disordered Weyl semimetals and 
coherent internodal tunnelling. 
We demonstrate that the gap $2\Delta$, created in the quasiparticle dispersion by such tunnelling, significantly affects
the longitudinal conduction (conduction along the magnetic field) of the quasiparticles if the magnetic field is perpendicular to the line connecting a pair of nodes in momentum space.
Quasiparticle states are strongly hybridised between the nodes in this regime, and their scattering is affected by long-range correlated disorder,
with the correlation length longer than the inverse separation between the nodes in momentum space.
The longitudinal resistivity in this regime depends exponentially on the magnitude of the magnetic field.

For the other directions of the magnetic field, the internodal hybridisation of the quasiparticle states may be neglected. The resistivity
is then determined by large-momentum scattering
between states at different nodes and has a strong dependence on the direction of the field.

Our results demonstrate how the internodal coupling (the gaps in the spectra of Landau levels) in Weyl semimetals may be observed in transport experiments.
Furthermore, the strong dependence of the internodal coupling on the external magnetic field can be used in valleytronic devices
to control electron transport if the magnetic field is perpendicular to the separation between the nodes.

The paper is organised as follows. In Sec.~\ref{Sec:results} we summarise our results for the
magnetoconductance of a Weyl semimetal in a magnetic field. A detailed microscopic 
discussion of the model of a disordered two-node Weyl semimetal and its typical parameters
is presented in Sec.~\ref{Sec:Model}. We discuss the quasiparticle dispersion in the system
in Sec.~\ref{Sec.Quasiparticle_disperion-in_a_disorder-free_semimetal}. 
Sections~\ref{Sec:Magnetoconductivity_away_from_eta=pi/2} and \ref{Sec:Magnetoconductivity_at_eta=pi/2} deal with quasiparticle scattering off
impurities and magnetoconductance in disordered semimetals. In Sec.~\ref{Sec:Magnetoconductivity_away_from_eta=pi/2} we consider
generic directions of the magnetic field where the magnetoresistance is weakly affected 
by the internodal coupling. Finally, in Sec.~\ref{Sec:Magnetoconductivity_at_eta=pi/2} we
demonstrate that the conduction is strongly affected by the coupling if the magnetic field is perpendicular 
to the line connecting the nodes and compute the resistivity in this regime.


\section{Results}

\label{Sec:results}

In the presence of an external magnetic field in a sufficiently clean system, the quasiparticle motion is quantised in the plane perpendicular to the field.
At the zeroth Landau level, quasiparticles at each Weyl nodes
may move only in one direction,  either parallel or antiparallel to the magnetic field, depending on the chirality of the node,
while quasiparticles at higher Landau levels may move both parallel and 
antiparallel to the magnetic field.

As a result, at strong magnetic 
fields or low levels of doping, for which only the zeroth Landau level contributes to transport (the {\it ultraquantum limit}),
the longitudinal resistivity of a Weyl semimetal is determined by the internodal scattering.
There are two main mechanisms of such internodal scattering: 1) large-momentum scattering between states at different nodes by impurities and
2) the interplay of small-momentum processes near one node and the internodal hybridisation of quasiparticle states.

In this paper, we analyse the dependence of the magnetoconductivity in a Weyl semimetal with two nodes
as a function of the magnetic field $B$ and its direction,
 focussing on the interplay of the internodal tunnelling and 
magnetoresistance. In the ultraquantum limit, 
the longitudinal resistivity of a Weyl semimetal with charged impurities
may be described by the interpolation formula
\begin{align}
\rho(\eta,\bB)=\rho_{\text{inter}}\cos^2\eta + \rho_{\text{gap}}(B,\eta),
\label{ResistivityContributions}
\end{align}
where
$\rho_{\text{inter}}\cos^2\eta$ is the contribution of large-momentum elastic scattering and is weakly dependent on the
absolute value $B$ of the magnetic field;
the last term $\rho_{\text{gap}}(B,\eta)$ accounts for the effects of the gap in the quasiparticle dispersion at the zeroth Landau level 
caused by coherent internodal tunnelling; $\eta$
is the angle between the field 
and the separation between the nodes in momentum space.

For screened charged impurities, ubiquitous in Weyl and semiconducting systems, the function $\rho_{\text{inter}}$ weakly
depends on the absolute value $B$ of the magnetic field
in the experimentally important limit $l_B Q\gg1$, where $l_B$ is the magnetic length and $2Q$ is the momentum separation between the nodes. 
Magnetoresistance independent of the absolute value of the magnetic field has also been demonstrated previously
(see, e.g., Refs.~\onlinecite{LuSheng:review,Goswami:Magnetoresistance,BehrendsBardarson:strainConductance}) for the special case of Gaussian impurities of the width $a\ll l_B$.
For the magnetic fields corresponding to $l_B Q\gg 1$ (the limit opposite to the one considered here),
 a Weyl semimetal with charged impurities exhibits
a different dependence
of the resistivity ($\rho_{\text{inter}}\propto 1/B^2$) on the magnitude of the magnetic field~\cite{Goswami:Magnetoresistance}.

In this paper, we also find a strong angular dependence $\propto \cos^2\eta$ of the first contribution $\rho_{\text{inter}}$
to the resistivity \eqref{ResistivityContributions}. Such anisotropic behaviour of the resistivity comes the 
structure of the wavefunctions at the nodes.

The second contribution, $\rho_{\text{gap}}(B,\eta)$ is suppressed in the experimentally important case of weak magnetic fields and is, therefore,
important at angles close to $\eta=\pi/2$. This contribution exhibits an exponential dependence
\begin{align}
\rho_{\text{gap}}(B,\pi/2)\propto\exp\left(-B_0/B\right)
\end{align}
on the magnitude of the magnetic field, where the characteristic field $B_0$ is given by the integral 
\begin{align}
B_0=\frac{2c}{|e|v_F \hbar}\int_\text{node 1}^\text{node 2} \xi_\bk dk
\end{align}
of the quasiparticle dispersion $\xi_\bk$ along the minimum-action tunnelling path between the nodes in momentum space;
$v_F$ is the Fermi velocity which is assumed to be isotropic near each node. 
The exponential dependence of the resistance \eqref{ResistivityContributions} on the magnetic field comes 
from the existence of the gap~\cite{SaykinTikhonovRodionov} in the dispersion of the quasiparticles, where the characteristic field
$B_0$ may be on the order of $10T$ or larger~\cite{SaykinTikhonovRodionov}.

Away from the ultraquantum limit, i.e.
at higher levels of doping or smaller magnetic fields, non-zero Landau levels contribute to the resistivity.
Unlike the zeroth Landau level, quasiparticles at those higher Landau levels can be backscattered within the same node.
Because of the long-range-correlated nature of the potential of screened impurities,
such intranodal scattering at higher Landau levels is significantly stronger than the internodal scattering.
The resistivity in this regime has recently been a subject of numerous studies (see, e.g., Refs.~\onlinecite{SonSpivak:NMR,Lee:screeningBnotAccounted,Reis:longitudinalMRexp,Burkov:review}) and is not a focus of this paper.

At the levels of doping and magnetic fields where the higher Landau levels contribute to transport,
their contribution $\rho_{\text{intra}}(B)$ to the resistivity may be added to the contributions~\eqref{ResistivityContributions}
on which we focus in this paper. Different contributions may be separated from each other using their different dependencies
on the direction and the value of the magnetic field. In the case of isotropic dispersions at Weyl nodes, the contribution $\rho_{\text{intra}}(B)$ of the intranodal scattering at high Landau levels is isotropic, which may be used to separate it from
the first term in Eq.~\eqref{ResistivityContributions}. It also has a weaker than exponential dependence on the magnetic field,
which distinguishes it from the effects of the gap at the zeroth Landau level described by the second term in Eq.~\eqref{ResistivityContributions}. 
For simplicity, in this paper we focus on the ultraquantum limit, where the contribution
$\rho_{\text{intra}}(B)$ is absent and the full resistivity is, thus, described by Eq.~\eqref{ResistivityContributions}.

We compute both contributions to the resistivity~\eqref{ResistivityContributions} microscopically. We find that
for the experimentally important case of charged impurities and sufficiently small magnetic fields
\begin{align}
	\rho_{\text{inter}}=\frac{2\pi^3 ne^2}{v_F^2 Q^4 \hbar \varepsilon^2},
	\label{Conductivity_without_splitting}
\end{align}
where $n$ is the concentration of the impurities and $\varepsilon$ is the dielectric constant.
For the gap-dependent contribution, which for the experimentally important 
case of low fields is relevant only at $\eta\approx\frac{\pi}{2}$ (or, equivalently, $\eta\approx -\frac{\pi}{2}$) we obtain
\begin{align}
	\rho_{\text{gap}}\left(B,\frac{\pi}{2}\right)
	=C(B)\frac{nc|e|\hbar^2}{\mu_0^2 B_\mu^2}B\,\min\left[1,\left(\frac{B}{B_\mu}\right)^3\right]
	e^{-\frac{B_0}{B}},
	\label{RhoExponentialInterpolation}
\end{align}
where $C(B)$ is a dimensionless coefficient weakly dependent on the magnetic field and where we introduced the characteristic magnetic field
\begin{align}
	B_\mu=\left(\frac{2\pi}{9\alpha}\right)^\frac{1}{3}\frac{c\mu_0^2}{|e|v_F^2\hbar},
	\label{Bmu}
\end{align}
with $\alpha=\frac{e^2}{\varepsilon\hbar v_F}$ being the so-called fine-structure constant
in a Weyl semimetal
 and $\mu_0$ being the chemical potential
in the system at zero magnetic field (measured from the energy of the Weyl node).

For a semimetal with the chemical potential 
$\mu_0=10meV$ (corresponding to the concentration of dopants $n\sim 10^{15}cm^{-3}$) and the dielectric constant~\cite{JAYGERIN1977771, Madelung, Jenkins:CarrierDynamics, Buckeridge:TaAsDielectricConstant} $\varepsilon=10-50$, the characteristic magnetic field $B_\mu$ lies, respectively, in the interval
$B_\mu=0.2-0.4T$.
Fields of this order of magnitude also correspond to entering the ultraquantum limit for the characteristic chemical
potentials under consideration.
For such fields, the gap-dependent contribution $\rho_{\text{gap}}(B,\eta)$ is strongly suppressed,
due to its exponential dependence on the magnetic fields, except for selected directions of the field, corresponding to
$\eta=\pm \pi/2$. Thus, we expect our results for the contributions to Eq.~\ref{ResistivityContributions} to hold
in realistic Weyl semimetals for magnetic fields up to several tesla. 
For stronger fields, the resistivity is dominated by the gap-dependent contribution 
$\rho_{\text{gap}}(B,\eta)$ and has a strong exponential dependence on the magnetic field for all directions of the magnetic field
(except $\eta=0,\pi$, where the gap vanishes).

We emphasise that, although in most of this paper we consider a system with a single pair of Weyl nodes,
our results may easily be generalised to the case of a system with multiple pairs of nodes, e.g., $\text{TaAs}$ or $\text{TaP}$.
The conductivity in such a system is given by the sum $\sigma(\bB)=\sum_i\left[\rho_{\text{inter}}\cos^2\eta_i + \rho_{\text{gap}}(B,\eta_i)\right]^{-1}$ of contributions from pairs of close nodes, where $\eta_i$ is the angle between the field
and the difference of momenta of the nodes in the $i$-th pair.


\section{Model}

\label{Sec:Model}

We consider a Weyl semimetal (WSM) with two nodes separated along the $z$ axis in momentum space.
The Hamiltonian of the quasiparticles in this system, in the presence of charged impurities, is given by
\begin{align}
 \hat{H} =
   &v_F \left(\hat{k}_x - \frac{e}{c} A_x  \right) \hat\sigma_x 
  + v_F\left(\hat{k}_y   - \frac{e}{c}{A_y} \right) \hat\sigma_y 
  \nonumber\\ 
  &+  \hat\sigma_z\,\, m\left(\hat k_z - \frac{e}{c} A_z\right)
 + \frac{e^2}{\epsilon} \sum\limits_i Z_i\frac{e^{-\varkappa |\br - \bR_i|}}{ |\br - \bR_i|},
 \label{Hamiltonian}
\end{align}
\begin{figure}[h!]
	\centering
	\includegraphics[width=\linewidth]{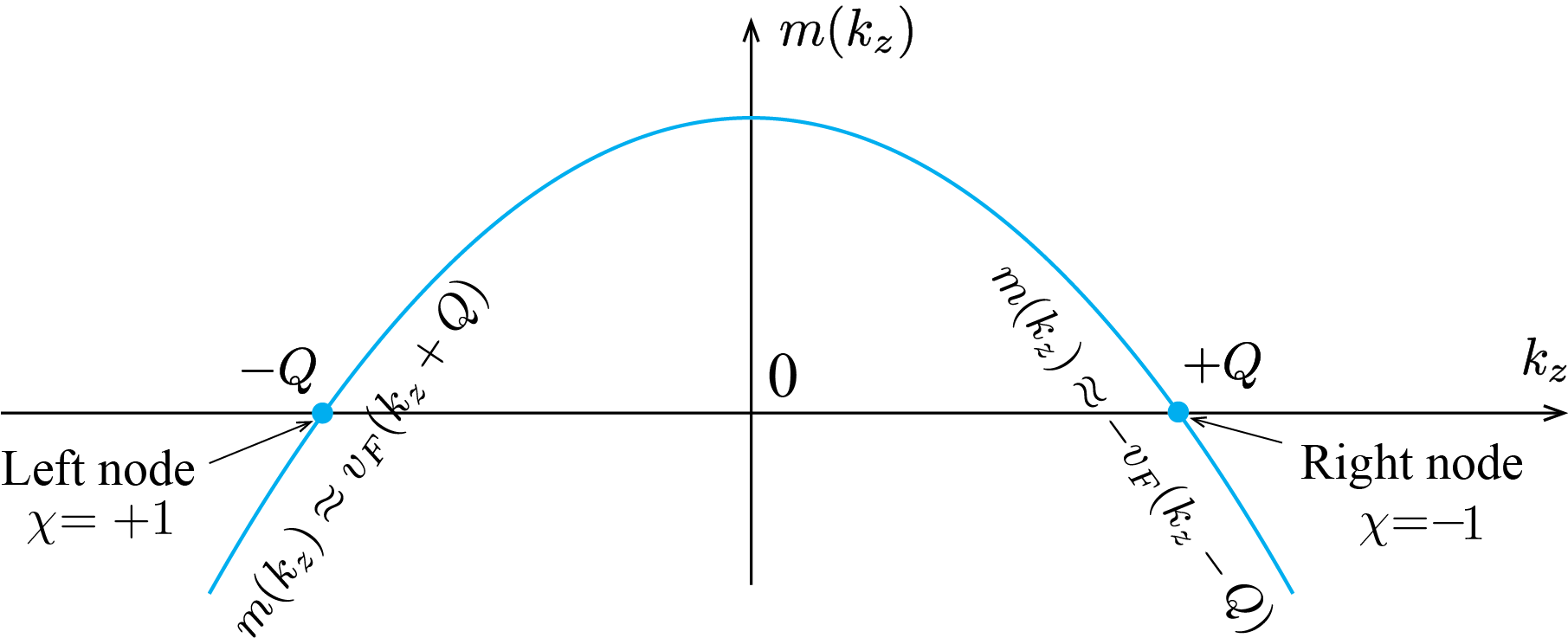}
	\caption{	\label{fig:m}
		(Colour online) The function $m(k_z)$ which determines the dispersion of the quasiparticles
		as a function of the momentum $k_z$ for $k_x=k_y=0$ [cf. Eq.~\eqref{Hamiltonian}].}
\end{figure}
where the first 
three terms in the right-hand side (rhs) represent the kinetic energy of the quasiparticles
in a magnetic field and the last term accounts
for the potential of screened impurities; $\hat\bk=(\hat k_x,\hat k_y,\hat k_z)=
-i\partial_\br$ is the operator of the quasiparticle 
momentum (hereinafter $\hbar=1$); $\hbsigma=(\hat\sigma_x,\hsigma_y,\hsigma_z)$ is the pseudospin operator, a degree of 
freedom equivalent to a spin-$1/2$; the vector potential $\bA=(A_x,A_y,A_z)$ accounts for the 
magnetic field; $\bR_i$ is the location of the $i$-th impurity; 
$Z_i e$ is the charge of the $i$-th impurity;
$\varkappa$ is the inverse screening length which is specified below;
$\varepsilon$ is the dielectric constant. For simplicity, we consider only two types of impurities:
with $Z_i=+1$ (donors) and with $Z_i=-1$ (acceptors).

In Eq.~\eqref{Hamiltonian} we introduced also a non-linear function $m(k_z)$ which
may be approximated by the linear dependencies 
$m(k_z)\approx  v_F \left(Q \mp k_z\right)$  
near the Weyl nodes, located at the momenta $\bk = (0, 0, \pm Q)$, as shown in Fig.~\ref{fig:m},
where the prefactor $v_F$ is assumed, for simplicity, to match the velocity of the quasiparticles
along the $x$ and $y$ axes in the absence of the magnetic field.
The free quasiparticles, thus, have linear (Weyl) dispersions near the nodes with the velocity $v_F$
and with opposite
chiralities~\footnote{The value of the chirality assigned to each node here matches the flux
$\chi=\frac{1}{2\pi}\int \cB(\bk) dS_\bk$ of the field $\cB=-i\bnabla_\bk\times\bra{\psi_+(\bk)}\bnabla_\bk\ket{\psi_+(\bk)}$ (measured in units of $2\pi$)
through a surface in momentum space 
surrounding the node,
where $\ket{\psi_+(\bk)}$ is the state of a quasiparticle with momentum $\bk$ in the conduction band of a disorder-free semimetal in zero magnetic field.}
 $\chi=\pm 1$ at the respective nodes.
The values of $m(k_z)$ away from the nodes determine the amplitude of quasiparticle tunnelling between the
nodes in momentum space. Such tunneling is usually neglected in studies of transport 
in WSMs. In this paper, we demonstrate, however, that the internodal tunnelling has a qualitative
effect on magnetotransport for certain directions of the magnetic field.




We choose the $xz$ plane to be parallel to the direction of the magnetic field,
as well as to the separation between the Weyl nodes in momentum space. In what follows, we use 
the Landau gauge 
\begin{align}
	\bA = (By \cos\eta, 0, - By \sin\eta),
	\label{VectorPotential}
\end{align}
 where $\eta$ is the angle between
the direction of the magnetic field $\bB$ and the $z$ axis, as shown in Fig.~\ref{Fig:Nodes_and_B}.

The magnetic field leads to the quantisation of the quasiparticle motion in the plane perpendicular
to the field in a sufficiently clean system. While we expect our results 
to hold for arbitrary magnetic fields, we consider, for simplicity, the {\it ultra-quantum limit},
where only the zeroth Landau level contributes to the transverse motion. This limit corresponds to 
sufficiently strong magnetic fields,
\begin{align}
	B>\frac{\mu^2 c}{2e^2 v_F^2},
\end{align}
where $\mu$ is the chemical potential measured from the energy of the Weyl nodes.
We assume also that $Q$ is the largest momentum scale in the problem and, in particular, exceeds the inverse magnetic length
\begin{align}
	\frac{1}{l_B}=\sqrt{\frac{|e|B}{c}}.
	\label{MagneticLength}
\end{align}

{\it Impurity potential.}
The last term in the Hamiltonian~\eqref{Hamiltonian} describes the potential of the screened impurities.
The inverse screening radius $\varkappa$ is given, in the Thomas-Fermi approximation and in the ultra-quantum limit under consideration,
by (see Appendix~\ref{Screening_of_impurities_due_to_the_zeroth_Landau_level} for details)  
\begin{eqnarray}
	\varkappa = \sqrt{\frac{2|e|^3 B}{\pi\epsilon v_F c}}\equiv\sqrt{\frac{2 \alpha}{\pi}} \frac{1}{l_B},
	\label{Screening_const}
\end{eqnarray}
where $l_B$ is the magnetic length, given by Eq.~\eqref{MagneticLength}, and $\alpha = \frac{e^2}{v_F \epsilon}$ is the ``fine-structure constant''.
Because the dielectric constant $\epsilon$ is large~\cite{JAYGERIN1977771, Madelung, Jenkins:CarrierDynamics, Buckeridge:TaAsDielectricConstant}
in most Weyl and Dirac materials, the fine-structure constant $\alpha$ may be assumed to be small, which justifies
using the Thomas-Fermi approximation when describing electrostatic screening in these systems (see Appendix~\ref{Screening_of_impurities_due_to_the_zeroth_Landau_level} for a more detailed discussion). 
A factor of $2$ in Eq.~\eqref{Screening_const} reflects the number of Weyl nodes in the system.

Donor and acceptor impurities affect the concentration of the quasiparticles and, thus, determine the
chemical potential in the system. 
In an uncompensated system, i.e. at $n_A\neq n_D$, the chemical potential $\mu$ at realistic 
dopant concentrations may be considered homogeneous, with the average value significantly exceeding
spatial fluctuations~\cite{RodionovSyzranov:donorsAcceptors}.
By contrast, in well compensated systems, with $\left|\frac{n_A-n_D}{n_A+n_D}\right|\ll \alpha^\frac{3}{2}$, the fluctuations of 
the screened potential of the impurities are significant and lead to the formation of electron and hole puddles~\cite{Skinner:puddles}~\cite{RodionovSyzranov:donorsAcceptors}. In this paper, we focus on the regime of an 
uncompensated system, as more experimentally relevant, with a sufficiently homogeneous chemical potential.

Charge neutrality requires that 
\begin{align}
	n_D-n_A=\frac{B|e|\mu}{2\pi^2 v_F c},
	\label{NeutralityCondition}
\end{align}
where $n_A$ and $n_D$ are the concentrations of the acceptor and donor impurities. The right-hand side
of Eq.~\eqref{NeutralityCondition} gives the concentration of quasiparticles measured from that 
in a disorder-free system with $\mu=0$, which is related to the chemical potential $\mu_0$ in the absence of the magnetic field
as $n_D-n_A=\frac{\mu_0^3}{3\pi^2 v_F^3}$. 
Because the density of states of the charge carriers is affected by the magnetic field and their concentration $n_D-n_A$ is field-independent,
the chemical potential $\mu$ depends on the magnetic field. According to Eq.~\eqref{NeutralityCondition},
\begin{align}
	\mu(B)=\frac{2c\mu_0^3}{3B|e|v_F^2}.
	\label{ChemicalPotentialAsAFunctionOfB}
\end{align}


\section{Quasiparticle disperion in a disorder-free semimetal}
\label{Sec.Quasiparticle_disperion-in_a_disorder-free_semimetal}

\begin{figure}
	\centering
	\includegraphics[width=\linewidth]{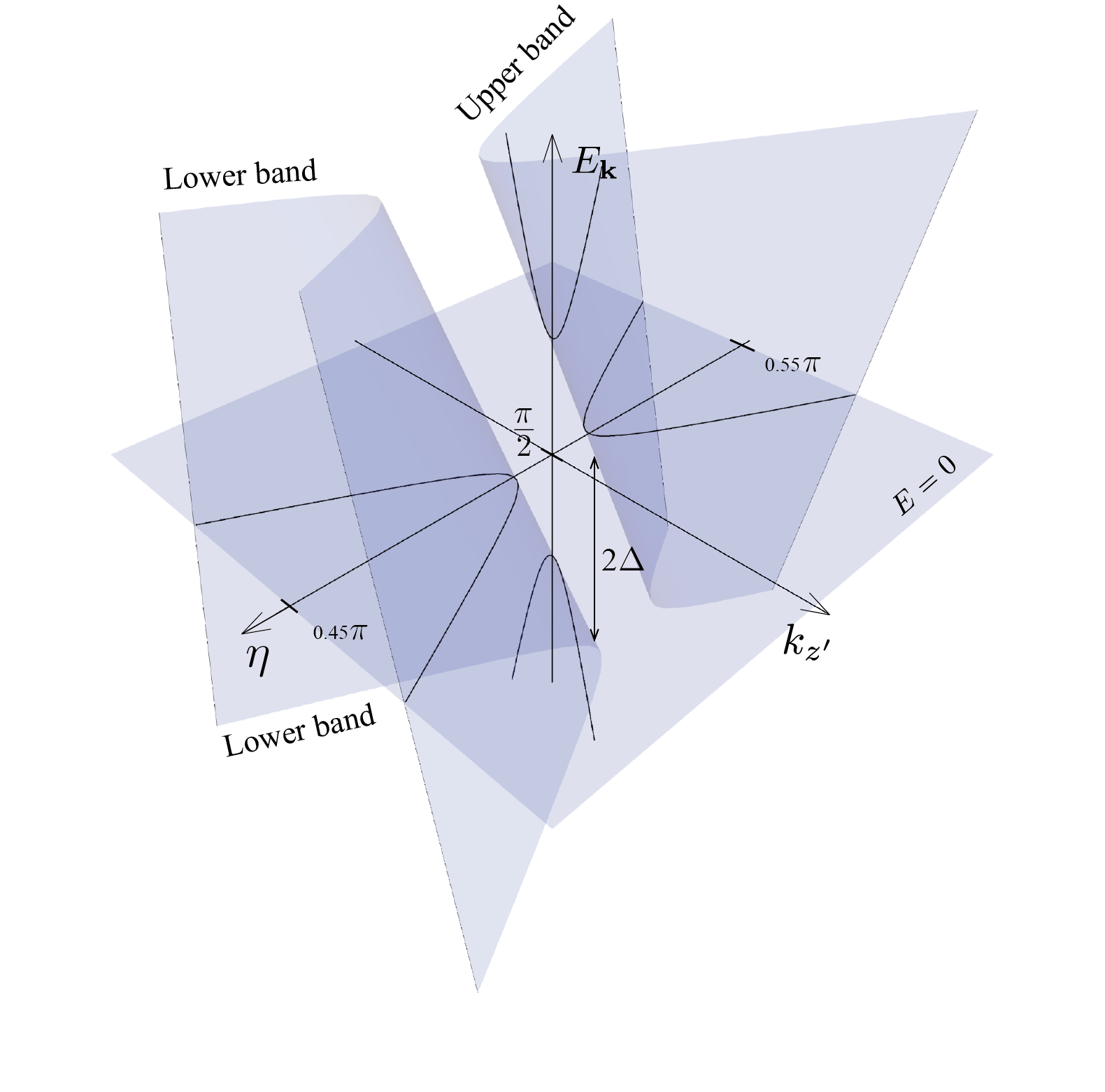}
	\caption{\label{fig:dispersionplot} (Colour online) The dispersion $E_\bk$ of the quasiparticles in a Weyl semimetal
		in a magnetic field at the zeroth Landau level as a function of the momentum $k_{z^\prime}$ along the magnetic 
	field and the angle $\eta$ between the field and the separation between the nodes for $\eta$ close to $\frac{\pi}{2}$.
	The dispersion has two bands, the ``upper band'' and the ``lower band''. The ``$E=0$'' plane corresponds to the 
	energies of the Weyl nodes in an undoped system.}	
\end{figure}

In this section, we analyse the dispersion of the quasiparticles at the zeroth Landau level
in a disorder-free system.
Figure~\ref{fig:dispersionplot} shows the dispersion, obtained by computing numerically the eigenenergies of the Hamiltonian~\eqref{Hamiltonian}
in the absence of impurities, 
as a function of the momentum $k_{z^\prime}$ along the magnetic field and the angle $\eta$ between the 
field and the separation between the nodes 
for $\eta$ close to $\pi/2$.

The dispersion consists of two bands (the ``upper band'' and ``lower band'' in Fig.~\ref{fig:dispersionplot})
separated by a gap $2\Delta$, which is determined by the internodal tunnelling of the quasiparticles.
For $\eta=\pi/2$, the dispersion is particle-hole symmetric, i.e. the two branches of the dispersion
are symmetric with respect to $E=0$. For angles $\eta$ deviating from $\pi/2$, the energies of both branches are shifted. 
At large momenta, $|k_{z^\prime}|\gg \Delta/v_F$, the dispersion is linear as a function of momentum.
In what follows, we describe the dispersion analytically.  

\subsection{Quasiparticle dispersion and wavefunctions for decoupled nodes ($\Delta=0$)}

First, we describe the dispersion 
and the wavefunctions of low-energy quasiparticles neglecting the internodal tunnelling.
This corresponds to either a negligible gap $2\Delta$, e.g., due to the magnetic field being small,
or to angles $\eta$ sufficiently away from $\pi/2$, where the
dispersion of low-energy quasiparticles (near the energy $E=0$) is described by the linear parts of the dispersion branches, 
independent of the quantity $\Delta$.

\begin{figure}
	\includegraphics[width=9cm,angle=0]{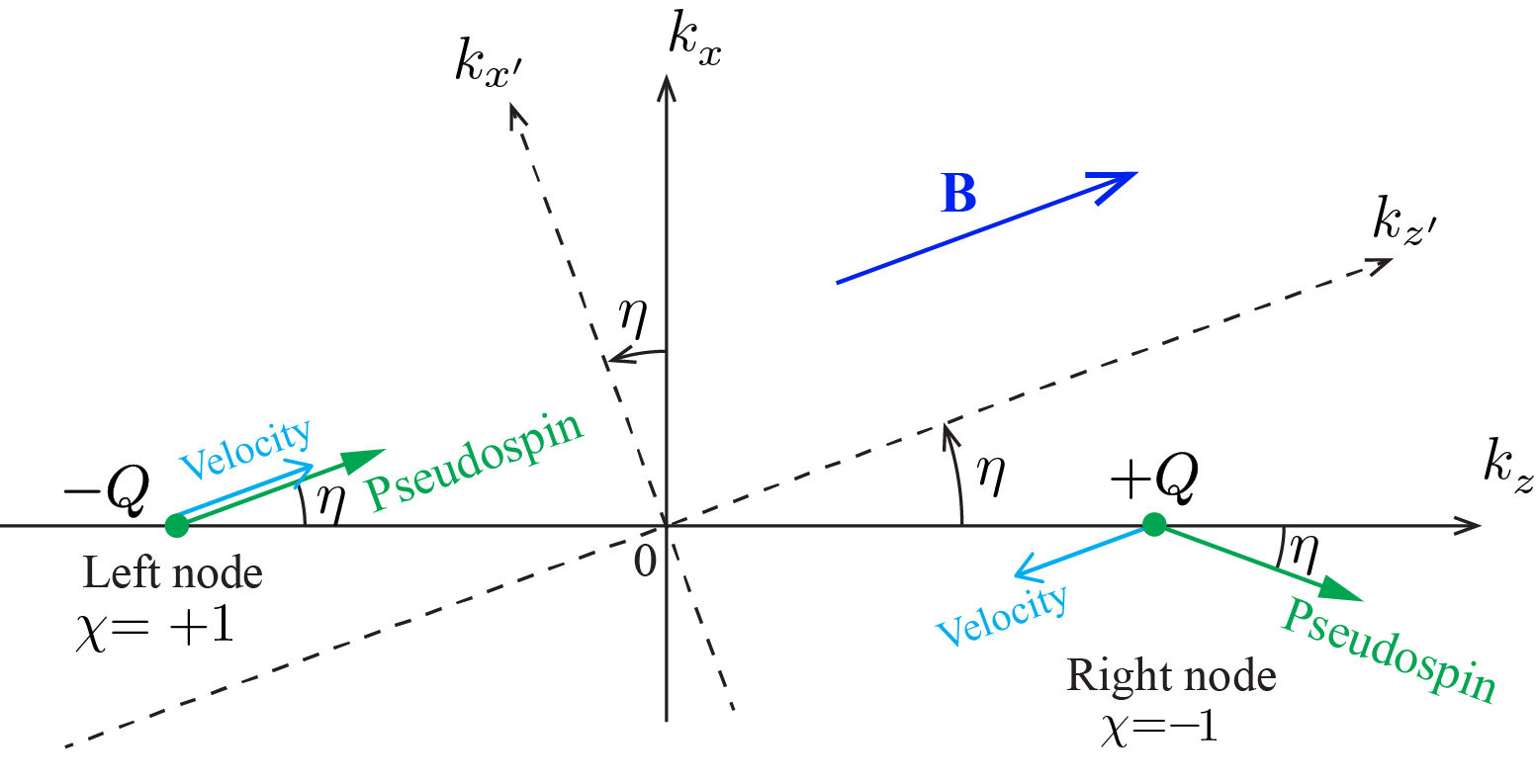}
	\caption{(Colour online)
		Weyl nodes and the directions of the magnetic field, quasiparticle velocities and the pseudospins 
		near the nodes.
		\label{Fig:Nodes_and_B}
	}
\end{figure}

We linearise the function $m(k_z)$ in the Hamiltonian~\eqref{Hamiltonian}
near the node of chirality $\chi$ (cf. Fig.~\ref{fig:m}),
\begin{align}
m(k_z) \approx v_F( Q + \chi k_{z}),
\label{m_expansion}
\end{align}
and rotate the coordinate frame by the angle $\eta$ about the $y$ axis, as shown in Fig.~\eqref{Fig:Nodes_and_B}. The $z^\prime$ axis of the rotated coordinate frame $x^\prime y z^\prime$
is parallel to the direction of the magnetic field $\bB$ and the components of the quasiparticle 
momenta in the $x^\prime z^\prime$ plane are given by
\begin{align}
	k_{x^\prime} = k_x \cos \eta - k_z \sin \eta
	\label{kxprime}
	\\
	k_{z^\prime} = k_x \sin \eta + k_z \cos \eta.
	\label{kzprime}
\end{align}
It is convenient to rotate also the pseudospin basis at each node of chirality $\chi$ by the
angle $\chi\eta$ about the $y$ axis in the pseudospin space, which corresponds to the transformation
of the Hamiltonian given by
\begin{align}
	\hat{H} \to \hat{H'}= e^{\frac{ i  \chi \eta\hsigma_y}{2} } \hat{H}  e^{- \frac{ i \chi \eta \hsigma_y}{2} }.
	\label{HamPseudospinRotation}
\end{align}

The Hamiltonian 
\begin{align}
	\hat{H}_0^\prime
	=
	&v_F\left(Q\cos\eta +\chi k_{z^\prime}\right)\hsigma_z
	\nonumber\\
	&+v_F\left(k_{x^\prime}-\chi Q\sin\eta-\frac{e}{c}By\right)-iv_F\partial_y
	\label{FreeHamiltonian}
\end{align} 
of free quasiparticles near the nodes in the rotated pseudospin basis may be rewritten conveniently, by introducing the annihilation
\begin{align}
	\ha = \frac{1}{l_B\sqrt{2}} \left( \partial_y +  k_{x^\prime} - \chi Q \sin \eta - \frac{e}{c}By \right),
	\label{aDefinition}
\end{align}
and creation $\ha^\dagger$ ladder operators of a harmonic oscillator,
in the form
\begin{align}
\hat{H}_0^\prime=
v_F\left(Q\cos\eta+\chi k_{z^\prime}\right)\hsigma_z + 
\frac{1}{\sqrt{2}}\frac{v_F}{l_B}
\hsigma^+ \ha^\dagger
+\frac{1}{\sqrt{2}}\frac{v_F}{l_B}
\hsigma^- \ha,
\label{HOneNode}
\end{align}
where $\hsigma^\pm=\hsigma_x\pm i\hsigma_y$.

The Hamiltonian~\eqref{HOneNode} with the ladder operator~\eqref{aDefinition} describes quasiparticles which propagate
parallel (or antiparallel) to the $z^\prime$ axis, i.e. along the direction of the magnetic field $\bB$, and whose motion
in the transverse direction is quantised.
The eigenstates of this Hamiltonian are parametrised by the component $k_{z^\prime}$ of momentum parallel to the magnetic field,
the number $n$ of the Landau level of the transverse motion and 
the transverse component $k_{x^\prime}$ of momentum. As discussed in Secs.~\ref{Sec:results} and \ref{Sec:Model}, we focus on the 
case of sufficiently strong magnetic fields, at which only the zeroth Landau level contributes to transport. 

The respective eigenstates of the Hamiltonian~\eqref{HOneNode} 
are given by
\begin{align}
	\ket{\psi_\chi} = \ket{\uparrow}_\chi\otimes\ket{0}_{LL}(k_{x^\prime})\otimes\ket{k_{z^\prime}},
	\label{PsiChi}
\end{align}
where $\ket{\uparrow}_\chi$ is the state of the pseudospin directed along the $z$ axis in the pseudospin space [where the pseudospin
basis depends on the chirality of the node $\chi$, cf.~\eqref{HamPseudospinRotation}], $\ket{0}_{LL}$ is the eigenstate of the 
operator $\ha$ corresponding to the zeroth Landau level  and parametrised by the momentum component $k_{x^\prime}$
 and $\ket{k_{z^\prime}}$ describes a plane-wave wavefunction of a particle with momentum
$k_{z^\prime}$ along the $z^\prime$ axis. In the coordinate representation and the pseudospin basis of the Hamiltonian~\eqref{Hamiltonian}
the eigenstate is given by
\begin{align}
\psi_{\chi\bk} (\br) =
e^{ - \frac{\left(  c k_{x^\prime} - c\chi Q \sin \eta- {e}By\right)^2}{2|e|B}+ik_{z^\prime}z^\prime+ik_{x^\prime}x^\prime }
\nonumber\\
\left( \frac{|e|B}{\pi cL_{z^\prime}^2L_{x^\prime}^2} \right)^{\frac{1}{4}}
\left(
\begin{array}{c}
\cos \frac{\eta}{2} \\  \chi \sin \frac{\eta}{2}
\end{array}
\right),
\label{LL_wavefunction}
\end{align}
where we assumed that the system has finite sizes $L_{z^\prime}$ and $L_{x^\prime}$ along the $z^\prime$ and $x^\prime$ axes;
$\bk$ is the momentum in the $xz$ plane (that matches the $x^\prime z^\prime$ plane).
The energy of this eigenstate is determined by the first term in the effective Hamiltonian~\eqref{HOneNode} and is given by
\begin{eqnarray}
E_{\chi\bk} = v_F(Q \cos \eta + \chi k_{z^\prime}).
\label{LL_energies}
\end{eqnarray}

The pseudospins of the low-energy quasiparticles under consideration have different directions, shown in Fig.~\ref{Fig:Nodes_and_B}, 
at nodes of different chiralities $\chi$. As follows from Eq.~\eqref{LL_energies}, the quasiparticle velocities
$\bv_{\chi\bk}=\frac{\partial E_{\chi\bk}}{\partial\bk}$
 at the nodes
with $\chi=+1$ and $\chi=-1$ are, respectively, parallel and antiparallel to the magnetic field, as shown in Fig.~\ref{Fig:Nodes_and_B}. 

\subsubsection{Generic wavefunction near a node}

\label{Sec:NearANode}

Although the Hamiltonian lacks translational invariance in the presence of the magnetic field, 
a generic wavefunction, considered as a function of the momentum $\bk$ in the $xz$ plane and the coordinate $y$,
may still be attributed to a particular node, so long as its generalised momentum 
$\bK=\bk-\frac{e}{c}\bA$ is close to the momentum of this node in the absence of the magnetic field,
 where $\bA(y)$ is the vector potential given 
by Eq.~\eqref{VectorPotential}.

Indeed, this condition is equivalent to 
1) the component $k_{z^\prime}$ of the momentum being close to $-\chi Q\cos\eta$ and 2) 
the argument $k_z-\frac{e}{c}A_z$ of the function $m$ in the Hamiltonian~\eqref{Hamiltonian} being close to $-\chi Q$.
The first condition guarantees that the wavefunction is a superposition of eigenstates with low energies (measured from
the energy of a node), while the second condition allows us to linearise the function $m(k_z)$ according to Eq.~\eqref{m_expansion}.


\subsection{Internodal tunnelling}
\label{Sec:Internodal_coupling}

The Hamiltonian~\eqref{Hamiltonian} describes quasiparticles with a finite bandwidth along the $z$ axis, which allows for quasiparticle
tunnelling from one node to the other. 
Although this tunnelling is exponentially suppressed by this bandwidth, which exceeds all the other energy scales in the problem, 
it leads to the opening of the gap $2\Delta$, shown in Fig.~\ref{fig:dispersionplot}, in the quasiparticle dispersion.
In this subsection, we consider analytically the effect of such internodal tunnelling in momentum space
on the quasiparticle dispersion and the opening of the gap.

The tunnelling leads to the hybridisation of low-energy quasiparticle states near different nodes.
The generic wavefunction $\psi(\br)$ of a quasiparticle with momentum components $k_x$ and $k_z$ in the $xz$ plane and with a sufficiently
low
energy (measured from the energy of the nodes) may be approximated as a
superposition of the wavefunctions \eqref{LL_wavefunction} belonging to the nodes with $\chi=+1$ and $\chi=-1$.

\subsubsection*{Conditions for strong internodal hybridisation}

Quasiparticle states at different nodes with the same momentum $k_{z^\prime}$ are strongly hybridised by the internodal
tunnelling if these states have the same energies in the absence of the tunnelling. Two quasiparticles states near the nodes 
with $\chi=+1$ and $\chi=-1$ have close energies~\eqref{LL_energies} if the momentum $k_{z^\prime}$ is sufficiently small.
Because the energies under consideration are also small, this leads to the requirement of the smallness of the momentum
scale $Q\cos\eta$ in comparison with the separation $Q$ between the nodes in momentum space and the characteristic momentum scale 
associated with the internodal tunnelling discussed below.

Therefore, the states at different nodes may be strongly hybridised only for angles $\eta$ sufficiently close to $\pi/2$,
\begin{align}
	\left|\eta-\frac{\pi}{2}\right|\ll\frac{\Delta}{v_F Q},
	\label{HybridisationCondition1}
\end{align}
where $\Delta$ is the characteristic coupling energy between the states with $\chi=+1$ and $\chi=-1$.

Strong hybridisation requires not only that the angle $\eta$ be close to $\pi/2$, but also that the energies of the hybridised states
be close to each other. At $\eta=\pi/2$, the energies of the states at nodes $\chi=+1$ and $\chi=-1$ are given
by $E_{+1\bk}=-E_{-1\bk}=v_F k_{z^\prime}=v_F k_x$, according to Eq.~\eqref{LL_energies}.
Therefore, strong hybridisation of states between nodes also requires that
\begin{align}
	v_F |k_x|\ll \Delta,
\end{align}
in addition to the condition~\eqref{HybridisationCondition1}.

\subsubsection*{Internodal coupling}

At angles $\eta$ very close to $\pi/2$, i.e. for magnetic fields almost perpendicular to the separation between the nodes,
the nodes are strongly coupled by the tunnelling and  
the effective Hamiltonian of low-energy quasiparticles is given by 
\begin{align}
	\cH_{\text{internodal}}(\bk)=
	&v_F\left(Q\cos\eta+k_{z^\prime}\right)\ket{\psi_{+1 \bk}}\bra{\psi_{+1 \bk}}
	\nonumber\\
	&+v_F\left(Q\cos\eta-k_{z^\prime}\right)\ket{\psi_{-1 \bk}}\bra{\psi_{-1 \bk}}
	\nonumber\\
	&+\Delta\left(\ket{\psi_{+1 \bk}}\bra{\psi_{-1 \bk}}+\ket{\psi_{-1 \bk}}\bra{\psi_{+1 \bk}}\right),
	\label{HamCoupled}
\end{align}
where $\bk$ is the momentum in the $xz$ plane, the states $\ket{\psi_{\chi \bk}}$ are given by Eqs.~\eqref{PsiChi} and 
\eqref{LL_wavefunction}, 
the first two lines describe the dispersion near decoupled nodes
and $\Delta$ determines the strength of the coupling. At $\eta=\frac{\pi}{2}$ the quantity $2\Delta$
gives the gap in the low-energy quasiparticle dispersion
\begin{eqnarray}
E_\bk = \pm \left({v_F^2 k_x^2 + \Delta^2}\right)^\frac{1}{2}
\label{Internodal_coupling}
\end{eqnarray}
shown in Fig.~\ref{fig:dispersionplot} (in the $\eta=\pi/2$ plane).

The amplitude of the tunnelling between the two nodes may be computed
(see Appendix~\ref{The_value_of_the_intervalle_coupling for_eta=pi/2})
 by matching the wavefunctions of the quasiparticles 
near the nodes with exponentially small ``tails'' between the nodes obtained in the quasiclassical approximation.
Recently, the internodal coupling for a particular form of the function $m(k_z)$ [cf. Eq.~\eqref{Hamiltonian}]
has been computed in Ref.~\onlinecite{SaykinTikhonovRodionov}.
In Appendix~\ref{The_value_of_the_intervalle_coupling for_eta=pi/2}, we generalise this calculation to the case of an arbitrary function
$m(k_z)$, with the result
\begin{align}
\Delta = v_F \sqrt{\frac{|e|B}{\pi c}} \exp\left[{ - \frac{c}{|e|B v_F} \int\limits_{-Q}^Q m(k_z) dk_z  }\right].
\label{Internodal_splitting_answer}
\end{align}
Equation \eqref{Internodal_splitting_answer} is valid for small values of the exponential, which reflects strong suppression 
of the quasiparticle wavefunctions away from the nodes. When deriving Eq.~\eqref{Internodal_splitting_answer} we also use the smallness
of the inverse magnetic length~\eqref{MagneticLength} compared to the separation $2Q$ between the nodes.


\section{Magnetoconductivity away from $\eta=\pi/2$}
\label{Sec:Magnetoconductivity_away_from_eta=pi/2}

In this section, we consider magnetotransport for angles $\eta$ away from $\pi/2$, at which the internodal tunnelling 
may be neglected and quasiparticles at different nodes are effectively decoupled. 
Because intranodal scattering processes do not change the velocities of the quasiparticles, shown in Fig.~\ref{Fig:Nodes_and_B},
the resistivity of the system in this regime is determined by the processes of internodal scattering.

As discussed in Sec.~\ref{Sec.Quasiparticle_disperion-in_a_disorder-free_semimetal}, the component $\bk$ of momentum
in the $xz$ plane is a good quantum number in a disorder-free system for the choice of the gauge of the vector potential given by
Eq.~\eqref{VectorPotential}.
Internodal scattering occurs between quasiparticle states $\psi_{\chi \bp}$ and $\psi_{-\chi\bk}$ with momenta $\bp$ and $\bk$
separated by a vector of approximate length $2Q$. While 
the difference between the momenta of quasiparticles near different nodes is evident
in the absence of magnetic field,
it requires a clarification in the presence of a magnetic field due to the lack of translational invariance of the Hamiltonian.

Indeed,
the change of the generalised momentum $\bk-\frac{e}{c}\bA(y)$ of a scattered quasiparticle is close to $2Q$.
This leads to the difference between the incoming and outgoing momenta $\bp$ and $\bk$
being close to $2Q$ if the random potential is sufficiently smooth. The 
potential of the screened impurities is smooth at length scales shorter than the screening length $\varkappa^{-1}\sim l_B/\alpha^\frac{1}{2}$.
Therefore, the difference between the momenta $\bp$ and $\bk$ is close to $2Q$ so long as momentum scales
on the order of $\alpha^\frac{1}{2}/l_B\ll Q$
 are neglected.
 This conclusion may also be obtained by evaluating explicitly the matrix elements of scattering of the states of the form~\eqref{LL_wavefunction}
 on the potential of screened impurities (see Appendix~\ref{Internodal_scattering_rate_away_from_eta=pi/2} for details).
 
The scattering rate of the state $\psi_{\chi \bp}$ at a node with chirality $\chi$ to the other node, with chirality $-\chi$,
is given by  
\begin{align}
\frac{1}{\tau_{\chi\bp}} = 2\pi \int
\left<
\left| \bra{\psi_{\chi \bp}}  U \ket{\psi_{-\chi \bk}} \right|^2
\right>_{\text{dis}}
\delta(E_{\chi \bp} - E_{-\chi \bk})
\frac{S_{xz} d\bk}{(2\pi)^2},
\label{Scattering_time_def}
\end{align}
where $U$ is the potential~\footnote{Hereinafter, the potentials $U(\br)$ and $u(\br)$ are the potential energies of the 
quasiparticles and not electrostatic potentials.}
of randomly located screened impurities; $\langle\ldots\rangle_{\text{dis}}$ is our convention for the averaging
over the locations of the impurities; $E_{\chi\bp}$ is the energy of the state with momentum $\bp$ near node $\chi$; $S_{xz}$
is the cross-sectional area of the system in the $xz$ plane (which, for simplicity, is assumed to be constant along the $y$ axis).

The potential $U(\br)$ is given by the last term in the Hamiltonian~\eqref{Hamiltonian} and is a sum of the potentials
$u_{i}(\br)=Z_i \frac{e^2}{\epsilon} \frac{e^{-\varkappa |\br - \bR_i|}}{ |\br - \bR_i|}$
of individual impurities at locations $\bR_i$. In this paper, we neglect single-particle interference effects related 
to scattering off multiple impurities~\cite{Abrikosov:metals,Gantmakher:book}, assuming that the concentration of impurities is small.

Under this approximation, we may make the replacement
\begin{align}
&\left<
\left| \bra{\psi_{\chi \bp}}  U(\br) \ket{\psi_{-\chi \bk}} \right|^2
\right>_{\text{dis}}
\nonumber\\
&\rightarrow
\sum\limits_i \frac{1}{V}\int d\bR_i\,  | \langle \psi_{\chi \bp} |   u(\br - \bR_i) | \psi_{-\chi \bk}  \rangle |^2
\label{Disorder_averaging}
\end{align}
in the expression~\eqref{Scattering_time_def} for the scattering time.
Using Eq.~\eqref{Disorder_averaging} and evaluating the integral in Eq.~\eqref{Scattering_time_def}
(see Appendix~\ref{Internodal_scattering_rate_away_from_eta=pi/2} for the details), 
we arrive at the internodal scattering rate
\begin{align}
\frac{1}{\tau} 
\approx
\frac{2\pi n |e|B}{v_F c} \left( \frac{e^2 \cos \eta}{2\varepsilon Q^2} \right)^2,
\label{Scattering_time_ans}
\end{align}
which is independent of the momentum of the scattered quasiparticle,
where $n=N/V$ is the concentration of the impurities in the system. 

The dependence $\propto\cos^2\eta$ of the scattering rate on the angle $\eta$ reflects the projection
of the pseudospins of quasiparticles at different nodes, shown in Fig.~\ref{Fig:Nodes_and_B}, onto each other.
Equation \eqref{Scattering_time_ans} suggests that the internodal scattering vanishes at $\eta=\pi/2$
as the pseudospins at different nodes are opposite to each other if the magnetic field is perpendicular to the line
separating the nodes. We emphasise, however, that the result \eqref{Scattering_time_ans} for the scattering rate 
applies at angles $\eta$ which are not very close to $\pi/2$, at which the hybridisation between quasiparticle states at different nodes
may be neglected.

Because the quasiparticles in an impurity-free system move parallel to the magnetic field,
the longitudinal conductivity (along the magnetic field) of a weakly disordered system significantly exceeds its transverse conductivity. 
When computing the longitudinal conductivity, on which we focus in this paper, transport may, therefore, be considered to be
one-dimensional, with 
$N_\bot = \frac{|e|B}{2\pi c}$ transverse channels
per cross-sectional unit area.

Since conduction comes only from low-energy quasiparticles, with energies significantly exceeded by the bandwidth,
their distribution function $f\left(k_{x^\prime},k_{z^\prime}\br\right)$ in the space of the momentum $\bk=(k_{x^\prime},k_{z^\prime})$
in the $xz$ plane and 3D coordinates $\br$ is peaked sharply near the surfaces $k_{z^\prime}=-\chi Q\cos \eta$; 
$k_{x^\prime}=\chi Q\sin \eta+\frac{eB}{c}y$, corresponding to node $\chi$, according to the discussion in Sec.~\ref{Sec:NearANode}.
Because the dynamics of the quasiparticles is effectively one-dimensional and confined to the zeroth Landau level in the transverse
direction, it is convenient to introduce the distribution function
$F\left(k_{z^\prime},\br\right)=\int_\chi \frac{dk_{x^\prime}}{2\pi}f\left(k_{z^\prime},k_{x^\prime}\br\right)$
of the longitudinal momentum $k_{z^\prime}$ and coordinate near node $\chi$, where the integration with respect 
to the momentum $k_{x^\prime}$ is carried out near that node.

The dynamics of this distribution function is governed by the kinetic equation
\begin{align}
	\frac{\partial F_\chi\left(k_{z^\prime},\br\right)}{\partial t}+
	\chi v_F\frac{\partial F_\chi\left(k_{z^\prime},\br\right)}{\partial z^\prime}+
	e E_{z^\prime}\frac{\partial F_\chi\left(k_{z^\prime},\br\right)}{\partial k_{z^\prime}}
	\nonumber\\
	=\frac{ F_{-\chi}\left(-k_{z^\prime},\br\right)- F_\chi\left(k_{z^\prime},\br\right) }{\tau}
	\label{KineticEquation}
\end{align}
for momenta $k_{z^\prime}$ close to the value $k_{z^\prime}\approx \pm Q \cos\eta$
near a node of chirality $\chi$.
Equation \eqref{KineticEquation} reflects that elastic internodal scattering leads to the scattering of states with
momentum $k_{z^\prime}$ along the $z^\prime$ axis into the states with momentum $-k_{z^\prime}$, as follows from the conservation of energy~\eqref{LL_energies}.

The generic stationary solution of the kinetic equation \eqref{KineticEquation} in a homogeneous system 
in the presence of a small electric field $E_{z^\prime}$ is given by
\begin{align}
	F_\chi\left(k_{z^\prime}+eE_{z^\prime}\tau\right)=F_{-\chi}\left(-k_{z^\prime}\right),
\end{align}
which leads to the current in the $z^\prime$ direction
\begin{align}
	j=eN_\bot v_F\int \left[F_{+1}(k_{z^\prime})-F_{-1}(k_{z^\prime})\right]\frac{dk_{z^\prime}}{2\pi}
	=\frac{|e|^3v_F B\tau}{(2\pi)^2c}E_{z^\prime}.
	\label{CurrentDecoupledNodes}
\end{align}
Utilising Eqs.~\eqref{CurrentDecoupledNodes} and \eqref{Scattering_time_ans} and recovering the Planck's constant $\hbar$
we arrive at the conductivity 
\begin{align}
	\sigma_{\text{inter}}=\frac{v_F^2 Q^4 \hbar \varepsilon^2}{2\pi^3 ne^2}\frac{1}{\cos^2\eta},
	\label{sigmaAnisotropic}
\end{align}
which describes the first contribution to the resistivity in Eq.~\eqref{ResistivityContributions}
with $\rho_{\text{inter}}$ given by~\eqref{Conductivity_without_splitting}.

As discussed in Sec.~\ref{Sec:results}, the conductivity \eqref{sigmaAnisotropic} is weakly dependent on the {magnitude of the} magnetic field,
{but} exhibits strong anisotropy, i.e. strong dependence on the {orientation of} magnetic field {with respect to the separation between the Weyl nodes}.
A strong dependence of the conductivity on the direction of the magnetic field has also been noted
in Ref.~\onlinecite{BehrendsBardarson:anisotropys}, however, with a different dependence on the direction. 
We believe that the difference comes from assuming in Ref.~\onlinecite{BehrendsBardarson:anisotropys}, without a derivation
for a particular model, a certain structure of the kinetic energy and the disorder potential in the space of the nodal spin.
We also emphasise the absence of the dependence of the conductivity \eqref{sigmaAnisotropic}
on the amplitude of the magnetic field.


\section{Magnetoconductivity at angles $\eta$ close to $\pi/2$}
\label{Sec:Magnetoconductivity_at_eta=pi/2}

In this section, we consider magnetotransport at angles $\eta$ close to $\pi/2$, i.e. for the magnetic field $\bB$
parallel to the $x$ axis. In this case, quasiparticle states at different nodes are coupled strongly by the internodal tunnelling.
As discussed in Sec.~\ref{Sec:Magnetoconductivity_away_from_eta=pi/2}, the internodal coupling determines essentially the 
resistivity in this regime, as the conductivity~\eqref{sigmaAnisotropic} diverges at $\eta=\pi/2$.

We assume here that the chemical potential $\mu$ (measured from the energy of the Weyl nodes in the absence of the magnetic field)
is greater than the internodal coupling $\Delta$.
Indeed, the chemical potential $\mu$ is on the order of several dozen meV for typical doping levels in WSMs (see, e.g., Refs.~\onlinecite{Jeon:CdAsSTM,Skinner:puddles,Shi:chargeDWWSM}),
whereas the coupling is of order several meV for fields $B\sim 10T$ and decreases exponentially for smaller fields.


In this approximation, the quasiparticle states in the absence of disorder may be represented in the form
\begin{align}
	\ket{\phi_{\chi \bk}}
	\approx \ket{\psi_{\chi \bk}}
	 +\frac{\chi\Delta}{2v_F k_x}\ket{\psi_{-\chi\bk}},
	 \label{WeaklyHybridisedStates}
\end{align}
as follows from the effective Hamiltonian \eqref{HamCoupled} of the quasiparticles at $\eta=\pi/2$. 
The second term in Eq.~\eqref{WeaklyHybridisedStates} is small, due to the smallness of the parameter $\Delta/(k_xv_F)\ll 1$,
and accounts for the hybridisation between the states $\ket{\psi_{+1 \bk}}$ and $\ket{\psi_{-1 \bk}}$
due to internodal tunnelling, which is essential for the conductivity being finite at $\eta=\pi/2$.

Quasiparticles with the wavefunctions $\ket{\phi_{+1\bk}}$ and $\ket{\phi_{-1\bk}}$ move with the velocity $v_F$ (up to small corrections
 on the order
of $\Delta^2/\mu^2$), respectively, parallel 
and antiparallel to the $x$ axis. The resistivity of the system is, thus, determined by the processes of scattering 
between the states $\ket{\phi_{\chi\bk}}$ with different $\chi$.

The respective ``internodal'' scattering rate is given by Eq.~\eqref{Scattering_time_def} with the replacement
$\ket{\psi_{\chi \bk}}\rightarrow \ket{\phi_{\chi\bk}}$. Utilising Eqs.~\eqref{Scattering_time_def}, \eqref{Disorder_averaging}, \eqref{WeaklyHybridisedStates}
and \eqref{LL_energies}, we obtain the scattering rate (see Appendix~\ref{Internodal_scattering_rate_away_at_eta=pi/2} for details)
\begin{align}
\frac{1}{\tau} 
\approx 
&\frac{2\pi n v_F e^4}{\varepsilon^2 \left[4\mu^2(B)+\varkappa^2(B) v_F^2\right]}
\frac{\Delta^2}{\mu^2(B)}
\nonumber\\
&\left[1+I\left(l_BQ,l_B\sqrt{4\mu^2(B)+\varkappa^2 v_F^2}/v_F\right)\right],
\label{Scattering_time_Bx}
\end{align}
where the function $I(s,t)$ is given by
\begin{align}
	I\left(s, t\right)=\frac{t^2}{\pi}\int 
	\frac{\cos\left(2s y\right) e^{-\frac{z^2+y^2}{2}}}
	{\left(y^2+z^2+t^2\right)^2}dz dy
	\label{IntegralI}
\end{align}
and the chemical potential $\mu(B)$ is given by Eq.~\eqref{ChemicalPotentialAsAFunctionOfB}.
Depending on the values of the magnetic field and the chemical potential, the value of the function
$I\left(l_BQ,l_B\sqrt{4\mu^2(B)+\varkappa^2 v_F^2}/v_F\right)$ may vary from zero, in the limit
$Q l_B \max\left(\sqrt{\frac{2\alpha}{\pi}},\frac{2\mu l_B}{v_F}\right)\gg 1$, to unity,
in the limit $Q l_B \max\left(\sqrt{\frac{2\alpha}{\pi}},\frac{2\mu l_B}{v_F}\right)\ll 1$ (see Appendix~\ref{Internodal_scattering_rate_away_at_eta=pi/2} for details).

For the experimentally relevant case of chemical potentials $|\mu(B)|\gg \Delta$ exceeding the gap $\Delta$ in
quasiparticle dispersion, the quasiparticles move with velocities
$v_F$ and $-v_F$ along the $z^\prime$ axis and the conductivity of the system is given by the same expression 
\begin{align}
\sigma=\frac{|e|^3 v_FB}{(2\pi)^2c}\tau
\label{Sigma}
\end{align}
as in Sec.~\ref{Sec:Magnetoconductivity_away_from_eta=pi/2} [cf. Eq.~\eqref{CurrentDecoupledNodes}].
Equations \eqref{Scattering_time_Bx} and \eqref{Sigma} give the longitudinal conductivity of a WSM in a magnetic field 
perpendicular to the line connecting the nodes in the form
\begin{align}
	\sigma = \frac{\varepsilon^2 B\left(\mu^2 + \varkappa^2 v_F^2\right)}
			{(2\pi)^3n|e|c(1+I)}\frac{\mu^2}{\Delta^2},
\end{align}
where $I=I\left(l_BQ,l_B\sqrt{4\mu^2(B)+\varkappa^2 v_F^2}/v_F\right)$ and $\Delta$
is the gap in the dispersion of the quasiparticles at the zeroth Landau level given by
Eq.~\eqref{Internodal_splitting_answer}.

In what immediately follows, we provide the results for conductivity in the two limiting 
cases: $\mu\gg \varkappa v_F$ and $\mu\ll\varkappa v_F$. According to Eqs.~\eqref{Screening_const}
and \eqref{ChemicalPotentialAsAFunctionOfB}, this corresponds, respectively, to the magnetic
fields $B\ll B_\mu$ and $B\gg B_\mu$, where the characteristic field $B_\mu$ is given by Eq.~\eqref{Bmu}
and is on the order of $0.1T$ for typical Weyl semimetals. For $B\ll B_\mu$, Eq.~\eqref{Sigma} gives
\begin{align}
	\sigma = \frac{2\varepsilon^2 c^4 \mu_0^{12}}{81\pi^2 ne^6v_F^{10}(1+I)}
			\frac{1}{B^4}
			\exp\left[{\frac{2c}{|e|B v_F} \int\limits_{-Q}^Q m(k_z) dk_z  }\right];
			\label{SigmaSmallB}
\end{align}
for $B\gg B_\mu$
\begin{align}
	\sigma = \frac{\varepsilon c\mu_0^6}{9\pi^3 n|e|v_F^5(1+I)}\frac{1}{B}
			 \exp\left[{\frac{2c}{|e|B v_F} \int\limits_{-Q}^Q m(k_z) dk_z  }\right].
			 \label{SigmaLargeB}
\end{align}
Combining Eqs.~\eqref{SigmaSmallB} and \eqref{SigmaLargeB} and recovering the Planck's constant $\hbar$
gives the interpolation formula \eqref{RhoExponentialInterpolation}.


\section{Summary and outlook}

In summary, we studied transport in a Weyl semimetal in a strong magnetic field in the presence of Coulomb impurities
and focussing on the case of a two-node Weyl semimetal.
The resistivity in the direction of the magnetic field 
exhibits a strong dependence, $\propto\cos\eta^2 + C$, on the angle $\eta$
between the direction of the magnetic field and the separation between the Weyl nodes in momentum space,
where $C\ll1$ is a small constant determined by the hybridisation of electron states between the nodes.

The strong directional dependence $\propto\cos^2\eta$ of the resistivity (away from the selected directions corresponding to $\eta=\pi/2$) of the system
along the magnetic field may
be compared directly with the results of experiments on magnetotransport in Weyl semimetals and used, in particular,
to identify new Weyl semimetals. At the same time, the resistivity $\rho(B)$ along the directions $\eta=\pi/2$ allows one to probe 
directly the internodal coupling $\Delta$, caused by the internodal tunnelling, which may affect phenomena involving 
internodal dynamics in Weyl semimetals, such as hydrodynamic electron flows~\cite{Gorbar:WeylHydro,Gorbar:hydroCS,LucasSachdev:WeylHydrodynamics,Yamamoto:HEPhydrodynamics,Galitski:dynamo,Gorbar:hydroNonLocal,Sukhachov:hydroCollective,Gooth:hydroObservation} or quantum oscillations of the resistance~\cite{Moll:SlabOscillations,Potter:OscillationsPrediction}, as well as valleytronic applications, if the magnetic 
field is perpendicular to the separation between the Weyl nodes.

We expect similarly anisotropic behaviour of the longitudinal resistivity in non-uniformly strained Weyl semimetals in the absence of 
the magnetic field, as elastic strain is known to generate gauge fields similar to those of the magnetic fields having different signs
for Weyl fermions of different chiralities~\cite{CortijoVozmediano:elasticGaugeFields,Pikulin:stranAnomaly,Cortijo:elasticStrain2}.
Internodal electron dynamics in a strained Weyl semimetal, however, requires a separate analysis, which we leave for future studies.


%
%

\section{Acknowledgments}

We are grateful to Ya.I.~Rodionov for useful discussions and previous collaboration on related topics.
Our work has been supported financially by the Hellman Foundation and the 
Faculty Research Grant awarded by the Committee on Research from the University of California
Santa Cruz (SVS) and by the Basic Research Programme of the HSE University and the RAS programme ``Low Temperature Physics'' (KST).


\onecolumngrid

\newpage

\appendix
\section{Screening of impurities in a strong magnetic field}
\label{Screening_of_impurities_due_to_the_zeroth_Landau_level}

In this section, we discuss the screening of a charged impurity
in a Weyl semimetal 
 in the presence of a magnetic field.
The electrostatic potential $\phi$ around an impurity of charge $Ze$ located at $\br=0$ is given by
\begin{align}
\varepsilon\,\Delta \phi(\br) = {4 \pi Z e }\, \delta(\br) + {4 \pi e n (\br)},
\label{Poisson_eq}
\end{align}
where $n(\br)$ is the change of the density of the electrons due to exposing the system to the potential $\phi$
at location $\br$. In the Thomas-Fermi approximation~\cite{Abrikosov:metals}, the response of the electron density $n(\br)$ to the potential $\phi(\br)$ is local,
which gives
\begin{align}
	n(\br)=N_{\text{nodes}}\frac{|e|B}{2\pi c}\int \frac{dk_z}{2\pi}\left\{n_F\left[v_F k_z-e\phi(\br)\right]-n_F\left(v_F k_z\right)\right\}
	=-N_{\text{nodes}}\frac{e^2 B \phi}{(2\pi^2) cv_F},
	\label{NTF}
\end{align}
where $\frac{|e|B}{2\pi c}$ accounts for the degeneracy of the Landau level; $v_F k_z$ is the dispersion of the quasiparticle along the magnetic field;
$N_{\text{nodes}}$ is the number of Weyl nodes in the system (in the rest of the paper, $N_{\text{nodes}}=2$) and $n_F(\varepsilon)$ is the Fermi-Dirac 
distribution function.

Equations \eqref{Poisson_eq} and \eqref{NTF} describe linear screening with the radius $\varkappa^{-1}$ which for $N_{\text{nodes}}=2$
is given by Eq.~\eqref{Screening_const}.
We note that the impurity screening may be neglected when considering the direct internodal scattering, because the change of momentum for such
scattering processes is close to $2Q$ and significantly exceeds the screening constant $\varkappa$.
Therefore, our results for the conductivity obtained in Sec.~\ref{Sec:Magnetoconductivity_away_from_eta=pi/2}, e.g., Eqs.~\eqref{Conductivity_without_splitting} and \eqref{sigmaAnisotropic}, are independent of the model of screening.

The screening and the details of the screened potential may be important, however, when considering the transport of quasiparticle states strongly hybridised between the nodes.
The respective contribution to the resistivity, considered in Sec.~\ref{Sec:Magnetoconductivity_at_eta=pi/2}, is determined by small momentum scattering 
on the order of $\min\left(|\mu|/v_F,\varkappa\right)$, where $\mu$ is the chemical potential in the system.

Thomas-Fermi approximation, which we used to obtain Eq.~\eqref{NTF} and the form of the screened potential given by the last term of the Hamiltonian~\eqref{Hamiltonian},
is justified provided the screening length $\varkappa^{-1}$ exceeds the characteristic lengths of the wavefunctions of the electrons which provide screening: the magnetic length $l_B$ and the wavelength $v_F/\mu$ corresponding to the motion along the field.
The condition $\varkappa^{-1}\gg l_B$ is fulfilled in typical Weyl semimetals due to the smallness of the ``fine structure constant''
discussed in Sec.~\ref{Sec:Model}. According to Eqs.~\eqref{Screening_const} and \eqref{ChemicalPotentialAsAFunctionOfB}, the condition  
\begin{align}
	\varkappa^{-1} \gg v_F/\mu
	\label{Condition:TFapplicability}
\end{align}
is fulfilled so long as the magnetic field $B$ is significantly smaller than the characteristic value 
given by Eq.~\eqref{Bmu}. The Thomas-Fermi approximation, used here, breaks down if the magnetic field exceeds this characteristic value,
which may affect the pre-exponential in Eq.~\eqref{SigmaLargeB} and in Eq.~\eqref{RhoExponentialInterpolation}
in the limit $B\gg B_\mu$, however, the results should still be expected to hold qualitatively.


\section{Landau levels of electrons near decoupled nodes}
\label{Sec:Landau_levels_without_internodal_coupling}

In this section, we present a detailed derivation of the wavefunctions of quasiparticle near each Weyl node neglecting the internodal 
tunnelling. The Hamiltonian of a disorder-free Weyl semimetal with two nodes and the quasiparticle dispersion linearised near the nodes is given by
\begin{align}
\hat{H}_0 = v_F \left( k_x - \frac{e}{c} By \cos\eta \right) \hsigma_x - i v_F  \hsigma_y \partial_y
+ v_F \left(Q + \chi k_z + \chi\frac{e}{c}By\sin\eta \right) \hsigma_z, 
\label{H0Appendix}
\end{align}
where the momenta $k_x$ and $k_z$ along the $x$ and $z$ axes are good quantum numbers. The eigenfunction $\psi_{k_x k_z\chi}$ of a quasiparticle at node $\chi$ and its eigenstate $E_\chi^2(k_x k_z)$ 
satisfies the equation 
$
	\hat{H}_0^2 \psi_{k_x k_z\chi} = E_\chi^2(k_x k_z)\psi_{k_x k_z\chi},
$
where the operator $\hat{H}_0^2$ is given by
\begin{align}
\hat{H}_0^2
= v_F^2 \left[
- \partial_y^2 + \left(k_x - \frac{e}{c} By \cos\eta\right)^2  + 
\left(Q + \chi k_z + \frac{e}{c}\chi By \sin\eta\right)^2
\right]
+
 \frac{e}{c}B v_F^2 \left(\hsigma_z\cos\eta +\chi\hsigma_x\sin\eta\right).
 \label{H02original}
\end{align}
After performing the 
transformation of the pseudospin basis given by Eq.~\eqref{HamPseudospinRotation} and the rotation of the coordinate frame given by
Eqs.~\eqref{kxprime} and \eqref{kzprime},
the operator $\left(\hat{H}_0^\prime\right)^2$, given by Eq.~\eqref{H02original}, takes the form
\begin{align}
\left(\hat{H}_0^\prime\right)^2
=
v_F^2
\left\{
-\partial_y^2  +  \left(k_{x^\prime} - \chi Q \sin \eta - \frac{e}{c} B y\right)^2
+ \left(k_{z^\prime} + \chi Q \cos\eta\right)^2  + \frac{e}{c} B \hsigma_z
\right\}.
\label{H02rotated}
\end{align}
The operators \eqref{H02original} and \eqref{H02rotated} have the form of the Hamiltonian of a system
consisting of a 
one-dimensional harmonic oscillator {with the mass $1/2v_F^2$ and the frequency $2v_F^2 |e|B/c$} and an independent spin-$1/2$ in a magnetic field.
To obtain the eigenstates of the system, it is convenient, therefore, to introduce the ladder operators
\begin{align}
&  \ha^{\dagger} =\sqrt{\frac{c}{2|e|B}}\left(  -\partial_y +  k_{x^\prime} - \chi Q \sin \eta - \frac{e}{c}By \right),
\label{acreation}
\\
& \ha = \sqrt{\frac{c}{2|e|B}} \left( \partial_y +  k_{x^\prime} - \chi Q \sin \eta - \frac{e}{c}By \right)
\label{aannihilation}
\end{align}
of this harmonic oscillator.

In terms of the creation and annihilation operators $\ha^\dagger$ and $\ha$, given by
Eqs.~\eqref{acreation} and \eqref{aannihilation}, the Hamiltonian
of the quasiparticles in the rotated pseudospin basis has the form  
\begin{align}
	\hat{H}_0^\prime=
	v_F\left(Q\cos\eta+\chi k_{z^\prime}\right)\hsigma_z + 
	\frac{1}{\sqrt{2}}\frac{v_F}{l_B}
	\hsigma^+ \ha^\dagger
	+\frac{1}{\sqrt{2}}\frac{v_F}{l_B}
	\hsigma^- \ha.
	\label{HamAlaOsc}
\end{align}
The eigenstates and the eigenenergies of the Hamiltonian \eqref{HamAlaOsc}, corresponding to the zeroth Landau level,
are given, respectively, by Eqs.~\eqref{PsiChi} and \eqref{LL_energies}.


\section{Intervalley coupling for $\eta=\pi/2$}
\label{The_value_of_the_intervalle_coupling for_eta=pi/2}

In this section, we present an explicit calculation of the internodal coupling $\Delta$ introduced in Sec. \ref{Sec:Internodal_coupling}. 
Because the coupling is essential at angles $\eta$ close to $\pi/2$, we focus on $\eta=\pi/2$, i.e. the direction of the magnetic
field perpendicular to the separation between the nodes. The coupling $\Delta$ for a particular form of the function $m(k_z)$ 
has been computed microscopically in Ref.~\onlinecite{SaykinTikhonovRodionov}. Here, we 
generalise this derivation to the case of a generic function $m(k_z)$.

Following Ref.~\onlinecite{SaykinTikhonovRodionov}, we perform a unitary rotation 
\begin{align}
	\psi \rightarrow \phi= e^{i\frac{\pi}{2}\hsigma_y}\psi
	\label{PhiPsi}
\end{align}
in the pseudospin space and rewrite
the equation $\cH^2\psi=E^2\psi$ 
for a disorder-free system in the form
\begin{align}
\left[
- v_F^2 \partial_y^2  +v_F^2 k_x^2 + m^2 \left(k_z + \frac{e}{c} By \right) - E^2
\right]
\phi
 = 
B m'\left(k_z + \frac{e}{c} By \right)
 \sigma_z
\phi.
\label{Schrodinger_eq}
\end{align}
We emphasise that near the node $\chi=-1$ the rotation~\eqref{PhiPsi} of the basis in the pseudospin space is distinct from that described by Eq.~\eqref{HamPseudospinRotation},
which we used in Sec.~\ref{Sec.Quasiparticle_disperion-in_a_disorder-free_semimetal} in order to obtain the quasiparticle wavefunctions in the absence of the
internodal coupling. Whereas the transformation~\eqref{HamPseudospinRotation} is different near different nodes, the rotation~\eqref{PhiPsi} is a global transformation of the basis;
the two transformations match near the node with $\chi=+1$ in the case $\eta=\pi/2$ under consideration.
In this section, we use also the coordinate frame described by Eqs.~\eqref{kxprime} and \eqref{kzprime} and shown in Fig.~\ref{Fig:Nodes_and_B} for $\eta=\pi/2$, i.e.
with $k_{z^\prime}=k_x$, $k_{x^\prime}=-k_z$.

Equation~\eqref{Schrodinger_eq} is similar to the Schr\"odinger equation of a quadratically dispersive particle
in a double-well potential~\cite{Landafshitz3}. The tunnelling amplitude between the two wells may be computed
in the quasiclassical approximation.
To that end, we introduce the classical (complex) momentum along the $y$ axis
\begin{align}
k_y (y) = \left[{ m^2 \left(k_z + \frac{e}{c} By\right) + v_F^2 k_x^2 - E^2 }\right]^\frac{1}{2}.
\end{align}
The quasiclassical solutions of Eq.~\eqref{Schrodinger_eq} in the region between the minima of the wells, where the amplitude
of the wavefunction is small, are given by 
\begin{align}
\phi_{1,2} = \sum_\pm C_{1,2}^{(\pm)} \frac{ e^{\pm \int\limits_{y_*}^y k_y dy \pm \int\limits_{y^*}^{y} \frac{dy eB m' }{2ck_y}}}{\sqrt{k_y}},
\label{Semiclassical_solution}
\end{align}
where $C_1^{(\pm)}$ and $C_2^{(\pm)}$ are constants and the coordinate $y^*$ may be chosen arbitrarily.
Between the nodes, the function $m$, which characterises the bandwidth of the quasiparticles, is large compared to 
all of the other momentum scales of low-energy quasiparticles. Therefore, in the region of applicability of Eq.~\eqref{Semiclassical_solution}
it is possible to make the approximation $k_y\approx m$.
Substituting the solutions \eqref{Semiclassical_solution} for the wavefunctions between the nodes into the Schr\"odinger equation $\hat H\psi=E\psi$ 
with the Hamiltonian~\eqref{Hamiltonian}
in the absence of the impurity potential, 
we obtain the relations 
\begin{align}
	C_2^{(\pm)} 
	= \left[\frac{(E\mp v_F k_x) }{2 m(y^{*})}\right]^{\pm 1}C_1^{(\pm)}
\end{align}
between the coefficients $C_1^{(\pm)}$ and $C_2^{(\pm)}$.

Equation~\eqref{Semiclassical_solution} describes the wavefunction between the nodes, where the exponential is small.
Solving the Schr\"odinger equation $\hat H\psi=E\psi$, with the function
$m$ linearised according to Eq.~(\ref{m_expansion}), gives the (non-normalised) wavefunction near a node of chirality $\chi$
in the form
\begin{align}
\left(
\begin{array}{c}
\phi_1  \\  \phi_2
\end{array}
\right)
= 
&\cos \left[
\frac{\pi c (E^2 - v_F^2 k_x^2)}{2v_F^2 eB}
\right]
e^{ - \frac{c \left(Q + \chi k_z +  \chi\frac{ e By}{c}\right)^2}{2|e|B} }
\left( Q + \chi k_z -  \chi\frac{ eBy}{c} \right)^{\frac{c(E^2 - v_F^2 k_x^2)}{2v_F^2 |e|B}}
\left(
\begin{array}{c}
1   \\
\frac{E- \chi k_x c}{2 \chi (Q c+ \chi k_x c - \chi By)}
\end{array}
\right)
\nonumber \\
&+\frac{\sqrt{2\pi}}{\Gamma \left[ \frac{c(E^2 - v_F^2 k_x^2)}{2v_F^2 eB} \right]  }
\left( \frac{|e|B}{2c} \right)^{\frac{c(E^2 - v_F^2 k_x^2)}{2v_F^2 |e|B} + \frac{1}{2}}
e^{  \frac{c \left(Q + \chi k_z +  \chi\frac{ e By}{c}\right)^2}{2|e|B} }
\left( Q + \chi k_z - \chi \frac{ eBy}{c} \right)^{\frac{c\left(E^2 - v_F^2 k_x^2\right)}{2v_F^2 |e|B} -1}
\left(
\begin{array}{c}
1 \\ \frac{2 \chi (Qc + \chi k_x c - \chi By)}{E+ \chi k_x}
\end{array}
\right),
\label{StraightSolutionNearOneNode}
\end{align}
where $\phi_1$ and $\phi_2$ are the components of the wavefunction $\phi$ with pseudospins, respectively, parallel and antiparallel to the $x$ axis.
The exponent $\frac{c \left(Q + \chi k_z +  \chi\frac{ e By}{c}\right)^2}{2|e|B}\equiv l_B^2 \left(\chi Q+\frac{eBy}{c}-k_{x^\prime}\right)^2$ in Eq.~\eqref{StraightSolutionNearOneNode} is small near the node of chirality $\chi$, in agreement with the discussion in Sec.~\ref{Sec:NearANode}.
As Eq.~\eqref{StraightSolutionNearOneNode} applies in the vicinity of the node, where the function $m$ may be linearised, it contains both factors contributions
exponentially decaying and exponentially increasing away from the node.

Matching the asymptotics of Eq.~\eqref{StraightSolutionNearOneNode} at large values of the quantity $Q+\chi k_z +\chi \frac{eBy}{c}$
with the solution~\eqref{Semiclassical_solution} between the nodes and obtaining the values of the coefficients $C_1$
and $C_2$, we arrive at the values of the quasiparticle energies in the form
\begin{eqnarray}
\frac{(E^2 - v_F^2 k_x^2)c}{2 |e| B v_F^2}
=
\frac{1}{2 \pi}
e^{ - \frac{2 c}{|e|B v_F} \int\limits_{-Q}^{Q}  m(p_z) dp_z},
\label{Delta_ans}
\end{eqnarray}
which leads to the value of the internodal coupling $\Delta$ given by Eq.~(\ref{Internodal_splitting_answer}) in the main text.


\subsection*{Weakly hybridised states}

At $\eta=\pi/2$, the dispersion of the quasiparticles at decoupled nodes is given by $E=\chi v_F k_x$, as follows from Eq.~\eqref{LL_energies}. 
The internodal coupling leads to the hybridisation of quasiparticle states at different nodes and the modification of the dispersion.
As discussed in Sec.~\ref{Sec:Internodal_coupling}, the states at $\eta=\pi/2$ get hybridised most strongly for small momenta $k_x$.

As we discuss also in 
Sec.~\ref{Sec:Magnetoconductivity_at_eta=pi/2}, the chemical potential $\mu$ in realistic WSMs corresponds to large 
momenta, $|k_x|\gg \Delta/v_F$, at which the states are weakly hybridised by the internodal tunnelling. 
Despite being weak, taking into account this hybridisation is essential for obtaining a finite conductivity in the case of the angle $\eta$
being close to $\pi/2$ we consider.

In the regime under consideration, a state with a positive momentum $k_x$ and
energy $E\approx v_F k_x + \frac{\Delta^2}{2v_F k_x}$ [cf. Eq.~\eqref{Internodal_coupling}] at the node $\chi=+1$ acquires, due to hybridisation, a small correction from the other node, $\chi=-1$.
The wavefunction of such a state is given by 
\begin{align}
\ket{\phi_{+1\bk}}
=
&\left( \frac{|e|B}{\pi c} \right)^{1/4}
e^{- \frac{c\left(Q+k_z + \frac{eBy}{c}\right)^2}{2|e|B}}
\left(
\begin{array}{c}
1  \\  0
\end{array}
\right)
\nonumber\\
\qquad
&- \sqrt{\frac{E- v_F k_x}{E + v_F k_x}}\left\{
\left( \frac{|e|B}{\pi c} \right)^{1/4}
e^{- \frac{c \left(Q-k_z- \frac{eBy}{c}\right)^2}{2|e|B}}
\left(
\begin{array}{c}
\frac{E + v_F k_x}{2m}  \\  1  
\end{array}
\right)
\right.
- 
\left.
\left( \frac{\pi c }{|e|B} \right)^{1/4}
\left(\frac{E}{v_F}+ k_x \right)
e^{ \frac{c\left(Q-k_z- \frac{eBy}{c}\right)^2}{2|e|B}}
\left(
\begin{array}{c}
1  \\  0
\end{array}
\right)
\right\}.
\label{Wavefunction_chi_p1}
\end{align}
The first term in the right-hand side of Eq.~\eqref{Wavefunction_chi_p1} describes the wavefunction of the quasiparticle
at node $\chi=1$. This term is peaked at $k_z=-Q-\frac{eBy}{c}$, i.e. at the location of the first node in the $xz$ plane in momentum space, as discussed in Sec.~\eqref{Sec:NearANode}.
The last two terms in Eq.~\eqref{Wavefunction_chi_p1} describe the small correction due to the presence of the other node, with the chirality $\chi=-1$. Those terms are peaked at $k_z=Q-\frac{eBy}{c}$
 and are suppressed by the small prefactor
$\sqrt{\frac{E- v_F k_x}{E + v_F k_x}}\approx \frac{\Delta}{vk_x\sqrt{2}}$ in the limit of large momenta $k_x\gg\Delta/v_F$
under consideration.

Similarly, the state at the node with chirality $\chi=-1$ is weakly hybridised, due  to the internodal tunnelling, with the other node
whose chirality is $\chi=-1$. The wavefunction of this state is given by
\begin{align}
	\ket{\phi_{-1\bk}}
	=
	&\left( \frac{|e|B}{\pi c} \right)^{1/4}
	e^{- \frac{c \left(Q-k_z-\frac{eBy}{c}\right)^2}{2|e|B}}
	\left(
	\begin{array}{c}
	0  \\  1
	\end{array}
	\right)
	\nonumber\\
	&- \sqrt{\frac{E + v_F k_x}{E - v_F k_x}}
	\left\{
	\left( \frac{|e|B}{\pi c} \right)^{1/4}
	e^{- \frac{c \left(Q+k_z+\frac{eBy}{c}\right)^2}{2|e|B}}
	\left(
	\begin{array}{c}
	1 \\  \frac{E - v_F k_x}{2m}    
	\end{array}
	\right)
	\right.
	- 
	\left.
	\left( \frac{\pi c}{|e|B} \right)^{1/4}
	\left(\frac{E}{v_F}- k_x \right)
	e^{ \frac{c\left(Q+k_z+\frac{eBy}{c}\right)^2}{2|e|B}}
	\left(
	\begin{array}{c}
	0  \\ 1
	\end{array}
	\right)
	\right\}.
	\label{Wavefunction_chi_p2}
\end{align}
The energy of this state is given by $E\approx -v_F k_x-\frac{\Delta^2}{2v_F k_x}$ for the large momenta $k_x$ under consideration.
The first and the last two terms in Eq.~\eqref{Wavefunction_chi_p2} describe, respectively, the state of a particle at the node
with chirality $\chi=-1$ and a correction due to the presence of the other node.


\section{Internodal scattering rate away from $\eta=\pi/2$}
\label{Internodal_scattering_rate_away_from_eta=pi/2}

In this section, we present details of the 
calculation of the internodal scattering rate~\eqref{Scattering_time_ans}
for angles $\eta$ away from $\pi/2$, where the effect of coherent internodal 
tunnelling on the quasiparticle dispersion and scattering may be neglected.
For simplicity, we assume that the system is rectangular, with the edges parallel to the $x^\prime$, $y$ and $z^\prime$ axes.

As discussed in Sec.~\ref{Sec:Magnetoconductivity_away_from_eta=pi/2}, the scattering rate is determined
by the matrix element of scattering of a quasiparticle near node $\chi$ with momentum $\bp=(\bp_{x^\prime},\bp_{z^\prime})$
in the $x^\prime z$ plane into the state with momentum $\bk=(\bk_{x^\prime},\bk_{z^\prime})$
near node $\chi^\prime$ on the potential $u(\br-\bR_i)$ of the $i$-th impurity, with the location
$\bR_i=\left(R_x^{(i)},R_y^{(i)},R_z^{(i)}\right)$ given by 
\begin{align}
\label{Matrix_element_expr}
\langle 
\psi_{\chi\bp} |  u(\br-\bR_i) | \psi_{\chi^\prime\bk}  
\rangle
\equiv
 \int d\br\, \psi^\dagger_{\chi\bp}  (\br) u(\br-\bR_i) \psi_{\chi^\prime\bk}  (\br) 
=\frac{4\pi Z_i  e^2}{V \varepsilon} \int d\br  \, \psi^\dagger_\chi (\br)  \psi_{\chi^\prime}(\br)
\sum \limits_{\bq}
 \frac{e^{i\bq\left(\br-\bR_i\right)}}{q^2 + \varkappa^2},
\end{align}
where $\psi_{\chi\bp}(\br)$ and $\psi_{\chi^\prime\bk}(\br)$ are the wavefunctions of the  
respective states, given by Eq.~\eqref{LL_wavefunction}.
In Eq.~\eqref{Matrix_element_expr} we used the Fourier transform $u(\bq)=\frac{4\pi Z_i e^2}{q^2+\varkappa^2}$ of the 
potential~\cite{Note2} of an impurity.

The conductivity is determined by internodal scattering, with $\chi=-\chi^\prime$, while intranodal
scattering, corresponding to $\chi=\chi^\prime$, has no effect on transport since it does not change the 
quasiparticle velocity. Using Eq.~\eqref{LL_wavefunction} for the states with $\chi=1$ and $\chi^\prime=-1$
gives the matrix element for internodal scattering in the form
\begin{align}
\left< 
\psi_{+1\bp} |  u(\br-\bR_i) | \psi_{-1\bk} 
\right> 
= \frac{ 4\pi Z_i e^2 \cos\eta}{V \varepsilon} \sum\limits_{q_y} 
\exp\left[{- \frac{c(p_{x'}-k_{x^\prime} - 2 Q \sin \eta)^2}{4|e|B} - \frac{cq_y^2}{4|e|B} + \frac{icq_y}{eB}\left( \frac{p_{x'}+k_{x^\prime}}{2} - \frac{eBR_y^{(i)}}{c}\right) }\right]
\nonumber\\
\frac{e^{ -i (p_{x'} - k_{x'}) R_{x'}^{(i)} - i (p_{z'} - k_{z'})R_{z'}^{(i)} }}{q_y^2 + (p_{x'} - k_{x'})^2 + (p_{z'} - k_{z'})^2 + \varkappa^2}.
\end{align}
Performing disorder averaging, $\DisAverage{\ldots}=\frac{1}{V}\int \ldots d\bR_i$, of the square 
of the matrix element over the location of the impurity gives
\begin{align}
\left<
\left| \bra{\psi_{+1 \bp}}   U \ket{\psi_{-1 \bk}}  \rangle \right|^2
\right>_{\text{dis}}
=
N  \left( \frac{4\pi e^2 \cos\eta }{V \varepsilon} \right)^2 e^{- \frac{c(p_{x'} - k_{x^\prime} - 2  Q \sin \eta)^2}{2|e|B}}
\sum\limits_{q_y}
\frac{e^{- \frac{c q_y^2}{2|e|B}} }{\left[ q_y^2 + (p_{x'} - k_{x'})^2 + (p_{z'} - k_{z'})^2 + \varkappa^2 \right]^2},
\label{Disorder_averaged_matrix_element}
\end{align}
where $N$ is the number of impurities in the system.

The expression in the denominator in Eq.~\eqref{Disorder_averaged_matrix_element} may be approximated as 
\begin{align}
	q_y^2 + (p_{x'} - k_{x'})^2 + (p_{z'} - k_{z'})^2 + \varkappa^2 \approx (2Q)^2.	
	\label{DenominatorApproximation}
\end{align}
Indeed, the sum with respect to $q_y$ in Eq.~\eqref{Disorder_averaged_matrix_element} is dominated by momenta $q_y$ on the order of the inverse 
magnetic length $l_B^{-1}$, which is given by Eq.~\eqref{MagneticLength} and is significantly smaller than the separation $2Q$
between the nodes.  
The inverse screening length $\varkappa$ is on the order of $\alpha^\frac{1}{2}/l_B$
and is also significantly exceeded by the momentum $Q$. 
Because the energies of the quasiparticles are small compared to the bandwidth and according to Eq.~\eqref{LL_energies},
$p_{z^\prime}\approx -Q \cos\eta$ and $k_{z^\prime}\approx Q\cos\eta$. The dynamics of the quasiparticles correspond to the range of momenta where
the function $m$ in the Hamiltonian~\eqref{Hamiltonian} may be linearised, Eq.~\eqref{m_expansion}. Together with
Eqs.~\eqref{kxprime} and \eqref{kzprime} this gives $k_{x^\prime}\approx Q\sin\eta$ and $p_{x^\prime}\approx -Q\sin\eta$.
Taking into account the values of all momenta and neglecting all momentum scales smaller than $Q$, we arrive at the approximation~\eqref{DenominatorApproximation}.

Performing the integration with respect to momentum $q_y$ (with the replacement
$\sum\limits_{q_y}\rightarrow L_y\int\frac{dq_y}{2\pi}$) and utilising Eq.~\ref{Disorder_averaged_matrix_element} gives
\begin{align}
\left<
\left| \bra{\psi_{+1 \bp}}   U \ket{\psi_{-1 \bk}}  \rangle \right|^2
\right>_{\text{dis}}
 \approx N   \left( \frac{4\pi e^2 \cos\eta }{V \varepsilon} \right)^2 e^{- \frac{c(p_{x'} - k_{x^\prime} - 2 Q \sin \eta)^2}{2|e|B}}
\frac{L_y}{2\pi} \sqrt{\frac{2|e|B\pi}{c}} \frac{1}{\left[ 4Q^2 \right]^2}.
\label{AvergedU}
\end{align}
Substituting the disorder-averaged scattering element \eqref{AvergedU} into the expression~\eqref{Scattering_time_def}
and performing integration over the momentum components $k_{x^\prime}$ and $k_{z^\prime}$, we
arrive at the expression~\eqref{Scattering_time_ans} for the internodal scattering rate.


\section{Internodal scattering rate at $\eta=\pi/2$}
\label{Internodal_scattering_rate_away_at_eta=pi/2}

In this section, we provide the details of the calculation of the quasiparticle scattering rate $1/\tau$ at 
$\eta=\pi/2$, i.e. for the magnetic field perpendicular to the separation between the nodes.
The resistivity at this angle is determined by the processes of scattering between states $\ket{\phi_{+1\bp}}$
and $\ket{\phi_{-1\bk}}$, given, respectively, by Eqs.~\eqref{Wavefunction_chi_p1} and \eqref{Wavefunction_chi_p2}.
The scattering rate between such states is given by
\begin{align}
\frac{1}{\tau} = 2\pi \int
\left<
\left| \bra{\phi_{+1 \bp}}  U \ket{\phi_{-1 \bk}} \right|^2
\right>_{\text{dis}}
\delta(E_{+1 \bp} - E_{-1 \bk})
\frac{S_{xz} d\bp}{(2\pi)^2},
\label{Scattering_time_def2}
\end{align}
where $E_{+1\bp}$ and $E_{-1\bk}$ are, respectively, the energies of the states $\ket{\phi_{+1\bp}}$ and $\ket{\phi_{-1\bk}}$
and $S_{xz}$ is the cross-section area of the system in the $xz$ plane.
%
%

Similarly to the case of internodal scattering at angles $\eta$ away from $\pi/2$, the scattering rate $1/\tau$ is determined by
the matrix element $\bra{\phi_{+1\bp}} |  u(\br-\bR_i) | \ket{\phi_{-1\bk}}$ of scattering off an impurity at location $\bR_i$.
Noticing that the exponentials in Eqs.~\eqref{Wavefunction_chi_p1} and \eqref{Wavefunction_chi_p2} which grow away from the 
nodes do not contribute to this matrix elements, we obtain
\begin{align}
&\bra{\phi_{+1\bp}} u(\br-\bR_i) \ket{\phi_{-1\bk}}=
\nonumber\\
&-  e^{-\frac{c(p_z - k_z)^2}{4|e|B}} \sqrt{\frac{E_{-1 \bk}+ v_F k_x}{E_{-1 \bk} - v_F k_x} } 
\frac{4 \pi Z_ie^2}{V\varepsilon}
\sum\limits_{q_y} \frac{e^{ - \frac{c q_y^2}{4|e|B} - \frac{ic q_y}{2eB}(2Q+p_z+k_z) - iq_y R_y^{(i)} - i (p_x-k_x)R_x^{(i)} - i (p_z-k_z)R_z^{(i)}}  }{q_y^2 + (p_x - k_x)^2 + (p_z - k_z)^2 + \varkappa^2}
\nonumber\\
&-  e^{-\frac{c(p_z - k_z)^2}{4|e|B}} \sqrt{\frac{E_{+1 \bp} -  v_F p_x}{E_{+1 \bp} + v_F p_x} } 
\frac{4 \pi Z_i e^2}{V\varepsilon}
\sum\limits_{q_y} \frac{e^{ - \frac{c q_y^2}{4|e|B} - \frac{ic q_y}{2eB}(-2Q+p_z+k_z) - iq_y R_y^{(i)} - i (p_x-k_x)R_x^{(i)} - i (p_z-k_z)R_z^{(i)}}  }{q_y^2 + (p_x - k_x)^2 + (p_z - k_z)^2 + \varkappa^2},
\label{MatrixElementPhi}
\end{align}
The first and the second lines in Eq.~\eqref{MatrixElementPhi} come from the wavefunctions near nodes $\chi=-1$ and $\chi=+1$.

The matrix element of scattering off the total potential $U(\br)=\sum_i u(\br-\bR_i)$, averaged over the 
realisations of the potential, is given by 
\begin{eqnarray}
\left<
|\bra{\phi_{+1 \bp}} U \ket{\phi_{-1 \bk}}|^2
\right>_{\text{dis}}
\label{Integral_of_matrix_element_Bx}
&&=
\frac{N}{V}\left<
|\bra{\phi_{+1 \bp}} u\left(\br-\bR_i\right) \ket{\phi_{-1 \bk}}|^2
\right>_{\text{dis}}
\nonumber\\
&&
= N \left( \frac{4 \pi e^2}{V \varepsilon} \right)^2 \frac{L_y}{2\pi}
\left\{
\left(  \frac{E_{-1\bk} + v_F k_x}{E_{-1\bk} - v_F k_x}  +  \frac{E_{+1\bp} - v_F  p_x}{E_{+1\bp} + v_F p_x}    \right)
\right.
\int \frac{dq_y\, e^{ -\frac{c(p_z - k_z)^2}{2|e|B} - \frac{c q_y^2}{2|e|B} } }{\left[  q_y^2 + (p_x - k_x)^2 + (p_z - k_z)^2 + \varkappa^2 \right]^2}
\nonumber\\
&& 
\qquad
+ 
\sqrt{\frac{(E_{-1\bk} + v_F k_x)(E_{+1\bp} - v_F p_x)}{(E_{-1\bk} - v_F k_x)(E_{+1\bp} + v_F p_x)}}
\left. \int \frac{dq_y \, e^{ -\frac{c(p_z - k_z)^2}{2|e|B} - \frac{c q_y^2}{2|e|B} }
	\left( e^{\frac{2iQc q_y}{eB}} +  e^{-\frac{2iQc q_y}{eB}}  \right)
}{\left[  q_y^2 + (p_x - k_x)^2 + (p_z - k_z)^2 + \varkappa^2 \right]^2}
\right\}.
\label{MatrElemTotalPotential}
\end{eqnarray}
The 
quantities $\left[  q_y^2 + (p_x - k_x)^2 + (p_z - k_z)^2 + \varkappa^2 \right]^\frac{1}{2}$ in Eq.~\eqref{MatrElemTotalPotential}
have the meaning of the effective momentum change of a plane-wave state scattered off the potential. We emphasise, however, that, unlike the case
of internodal scattering considered in Appendix~\ref{Internodal_scattering_rate_away_at_eta=pi/2}, this differences cannot be approximated by 
the separation $2Q$ between the Weyl nodes.

To compute the scattering time~\eqref{Scattering_time_def2}, we first perform the integration of the expression~\eqref{MatrElemTotalPotential} with
 respect to the momentum $p_z$. The first line gives a contribution proportional to 
\begin{align}
\int \frac{dp_z dq_y\, e^{-\frac{c(p_z - k_z)^2}{2|e|B}- \frac{c q_y^2}{2|e|B} }}{\left[ q_y^2 + (p_x - k_x)^2 + (p_z - k_z)^2 + \varkappa^2\right]^2}
= \frac{\pi}{(p_x -k_x)^2 + \varkappa^2}
- \frac{\pi c}{2|e|B}  e^{\frac{c(p_x -k_x)^2}{2|e|B} + \frac{c \varkappa^2}{2|e|B} }
E_1 \left[ \frac{c(p_x -k_x)^2}{2|e|B} + \frac{c\varkappa^2}{2|e|B}   \right],
\label{SomeIntegral1}
\end{align}
where $E_1(x)=\int_1^\infty dt\, e^{-tx}/t$ is the exponential integral. The characteristic momentum difference $p_x-k_x$ is on the order
of $|\mu|/v_F$, with $\mu$ being the chemical potential, and in the limit of strong magnetic fields under consideration (the ``ultraquantum'' limit) is exceeded
by the inverse magnetic length $\l_B^{-1}=\sqrt{|e|B/c}$. As a result and also considering the smallness of the magnetic length $l_B$
in comparison with the screening length $\varkappa$, the second term in the right-hand side of Eq.~\eqref{SomeIntegral1} may be neglected and we may approximate
\begin{eqnarray}
&&\int \frac{dp_z dq_y\, e^{-\frac{c(p_z - k_z)^2}{2|e|B}- \frac{c q_y^2}{2|e|B} }}{\left[ q_y^2 + (p_x - k_x)^2 + (p_z - k_z)^2 + \varkappa^2\right]^2}
\approx \frac{\pi}{(p_x -k_x)^2 + \varkappa^2}.
\label{Bx_integral_ans}
\end{eqnarray}

The integral of the second line of Eq.~\eqref{MatrElemTotalPotential} with respect to the momentum $p_z$ may be represented in the form
\begin{align}
	  \int \frac{dp_z dq_y\, e^{ -\frac{c(p_z - k_z)^2}{2|e|B} - \frac{c q_y^2}{2|e|B} }
		\left(  e^{\frac{2iQc q_y}{eB}} +  e^{-\frac{2iQc q_y}{eB}}  \right)
	}{\left[  q_y^2 + (p_x - k_x)^2 + (p_z - k_z)^2 + \varkappa^2 \right]^2}
	=  \frac{2\pi I\left[l_B Q, l_B \sqrt{(p_x-k_x)^2+\varkappa^2}\right]}{(p_x -k_x)^2 + \varkappa^2},
	\label{Bx_integral_ans2}
\end{align}
where the magnetic length $l_B$ is given by Eq.~\eqref{MagneticLength} and the integral $I(s,t)$ is given by Eq.~\eqref{IntegralI}.
Performing integration with respect to $p_z$ in the expression~\eqref{Scattering_time_def2} for the scattering rate and 
utilising Eqs.~\eqref{SomeIntegral1}, \eqref{Bx_integral_ans} and \eqref{Bx_integral_ans2}, we arrive at the scattering rate given by 
Eqs.~\eqref{Scattering_time_Bx} and \eqref{IntegralI}.

In the main text we consider the ranges of parameters that correspond to two limiting cases in the integral $I(s,t)$,
\begin{align}
I\left(s, t\right)=\frac{t^2}{\pi}\int 
\frac{\cos\left(2s y\right) e^{-\frac{z^2+y^2}{2}}}
{\left(y^2+z^2+t^2\right)^2}dz dy
\to
\begin{cases}
1, & st\ll 1, 
\\
0, & st\gg1.
\end{cases}
\end{align}


\twocolumngrid


%


\end{document}